\documentclass[useAMS,usenatbib]{mn2e}
\usepackage{aas_macros}
\usepackage{graphicx}
\usepackage{amsmath,amssymb}
\title[Brightest Cluster Galaxies in Cosmological Simulations]{Brightest Cluster Galaxies in Cosmological Simulations with Adaptive Mesh Refinement: Successes and Failures}
\author[D. Martizzi et al.]{\parbox[t]{\textwidth}{Davide Martizzi$^{1}$\thanks{E-mail: davide.martizzi@berkeley.edu}, Jimmy$^{2}$, 
Romain Teyssier$^{3}$, Ben Moore$^{3}$}\vspace*{6pt}\\
$^{1}$Department of Astronomy, University of California, Berkeley, CA 94720-3411, USA \\
$^{2}$George P. and Cynthia W. Mitchell Institute for Fundamental Physics and Astronomy, Department of Physics and Astronomy, \\
Texas A\&M University, College Station, TX 77843, USA \\
$^{3}$Institute for Computational Science, University of Zurich, CH-8057 Z\"urich, Switzerland\\
}

\begin{document}

%\pagerange{\pageref{firstpage}--\pageref{lastpage}} \pubyear{2010}
\maketitle

\label{firstpage}

\begin{abstract}

A large sample of cosmological hydrodynamical zoom-in simulations with Adaptive Mesh Refinement (AMR) is analysed to study the properties of simulated Brightest Cluster Galaxies (BCGs). Following the formation 
and evolution of BCGs requires modeling an entire galaxy cluster, because the BCG properties are largely influenced by the state of the gas in the cluster and by interactions and mergers with satellites. BCG 
evolution is also deeply influenced by the presence of gas heating sources such as Active Galactic Nuclei (AGNs) that prevent catastrophic cooling of large amounts of gas. We show that AGN feedback is one of the 
most important mechanisms in shaping the properties of BCGs at low redshift by analysing our statistical sample of simulations with and without AGN feedback. When AGN feedback is included BCG masses, sizes, star 
formation rates and kinematic properties are closer to those of the observed systems. Some small discrepancies are observed only for the most massive BCGs {and in the fraction of star-forming BCGs}, effects that might be due to physical processes that are not 
included in our model.

\end{abstract}


%%%%%%%%%%%%%%%%%%%%%%%%%%%%%%%%%%%%%%%%%%%%%%%%%%
\section{Introduction}

Brightest Cluster Galaxies (BCGs) are the most luminous and massive galaxies in the Universe. They typically sit at (or close to) the centre of the most massive gravitationally bound structures, galaxy clusters. 
BCGs are one of the best examples of hierarchical galaxy formation and evolution. Since they sit at the bottom of deep cluster potential wells they are exposed to a large number of events that modify their properties. 
Such processes include fast gas infall fueling star formation during their early-stage of evolution and dynamical interactions with satellite galaxies which fall into the cluster and eventually merge with the BCGs. On the 
other hand, the interaction of the satellite galaxies with their environment (e.g. other galaxies and the Intracluster Medium, ICM) determines the properties of the objects that merge with the BCGs, therefore influence 
the post-merging properties of the BCGs themselves. The stellar material stripped from the satellite galaxies during their interaction with the central potential ends up constituting the so-called Intracluster Light 
(ICL), i.e. an extended stellar halo surrounding the BCG and characterized by a rich variety of substructure \citep{Gonzalez:2005p907, Gonzalez:2007p5292, 2010ApJ...720..569R}. Metal enrichment and physical state of 
the ICM at low redshift determine whether the central region of clusters and BCGs will be characterized by cooling flows that feed star formation, i.e. the star formation rate of the central galaxy. BCGs also host 
the most most massive black holes in the Universe which can have masses up to a few $10^9$ M$_{\odot}$ \citep{2013ApJ...764..184M} and can influence the ICM state during their Active Galactic Nuclei (AGN) phases 
\citep{2012ARA&A..50..455F}. 

Given these considerations, it is appropriate to say that theoretical modeling of BCGs represents one of the most challenging problems in galaxy formation and evolution. Modeling BCGs requires also detailed modeling 
of the cluster environment and of the satellite galaxies. Ideal tools to study this problem are cosmological hydrodynamical simulations which try to model most of the relevant processes in galaxy formation. The problem 
of BCG formation has been already studied in detail in this context, e.g. by \cite{2013MNRAS.436.1750R}, and many results have been published on the properties of galaxy groups and clusters from the OWLs collaboration 
(e.g. \cite{2011MNRAS.412.1965M}) and other groups (e.g. \cite{Sijacki:2007p1032}). These studies have all been performed by analysing Smoothed Particle Hydrodynamics (SPH) simulations. Recently, we published the 
analysis of a series of high resolution cosmological hydrodynamical zoom-in simulations of a galaxy cluster and its BCG \citep{2012MNRAS.420.2859M} performed with the Adaptive Mesh Refinement (AMR) code {\scshape ramses} 
\citep{Teyssier:2002p451}. The novelty of the analysis of \cite{2012MNRAS.420.2859M} was that the BCG formation problem was addressed for the first time using an AMR code which included prescriptions for AGN feedback. 
In agreement with the result of other groups, we showed that realistic BCG properties could be obtained only if AGN feedback (or an equivalent source of heating) was included in the calculation to avoid catastrophic 
cooling of gas in the cluster core. Despite the excellent resolution of those calculations, the limit of the results shown in \cite{2012MNRAS.420.2859M} is that they are relative to one halo only. 

Studying the properties of BCGs provides a unique opportunity to test galaxy formation physics in AMR simulations. For this reason we performed a study complementary to that of \cite{2012MNRAS.420.2859M}. In this 
paper we analyse the BCGs in a large sample of cosmological hydrodynamical AMR zoom-in simulations (already presented in \cite{2014MNRAS.440.2290M}) at redshift $z=0$. The goal of this work is to study a statistical
sample of simulated BCG and place constraint on the validity of the adopted galaxy formation prescriptions.

The paper is organized as follows. Section 1 is dedicated to describing the methods and the simulations. Section 2 shows a detailed comparison of the simulated BCG properties to observational data. The final Section 
is a summary of the results and contains the main conclusions drawn from the analysis.

%%%%%%%%%%%%%%%%%%%%%%%%%%%%%%%%%%%%%%%%%%%%%%%%%%
\section{Numerical Simulations}
\label{sec:num_methods}

We consider a set of 102 cosmological re-simulations performed with the {\scshape ramses} code \citep{Teyssier:2002p451}. These simulations are part of a larger set recently used 
in \cite{2014MNRAS.440.2290M} to study the baryonic effects on the halo mass function. Thanks to the adaptive mesh refinement capability of the {\scshape ramses} code, the resolution 
achieved in these simulations is sufficient to study the properties of low redshift BCGs. The project required more than 3 million CPU hours.

\begin{table}
\begin{center}
{\bfseries Cosmological parameters}
\begin{tabular}{|c|c|c|c|c|c|}
\hline
\hline
 $H_0$ [km s$^{-1}$Mpc$^{-1}$] & $\sigma_{\rm 8}$ & $n_{\rm s} $ & $\Omega_\Lambda$ & $\Omega_{\rm m}$ & $\Omega_{\rm b}$ \\
\hline
\hline
 70.4 & 0.809 & 0.963 & 0.728 & 0.272 & 0.045 \\
\hline
\hline
\end{tabular}
\caption{Cosmological parameters adopted in our simulations. }\label{tab:cosm_par}
\end{center}
\end{table}

We follow the evolution of cosmic structure formation in the context of the standard $\Lambda$CDM cosmological scenario. In our calculations, the cosmological parameters are: matter density 
parameter $\Omega_{\rm m}=0.272$, cosmological constant density parameter $\Omega_\Lambda=0.728$, baryonic matter density parameter $\Omega_{\rm b}=0.045$, power spectrum 
normalization $\sigma_{\rm 8}=0.809$, primordial power spectrum index $n_{\rm s}=0.963 $ and Hubble constant $H_0=70.4$ km/s/Mpc (Table~\ref{tab:cosm_par}). The initial conditions for our 
simulations were computed using the \cite{Eisenstein:1998p1104} transfer function and the {\scshape grafic++} code developed by Doug Potter (http://sourceforge.net/projects/grafic/) and
based on the original {\scshape grafic} code \citep{Bertschinger:2001p1123}. 

First, we ran a dark matter only simulation with particle mass $m_{\rm cdm}=1.55\times 10^9$~M$_\odot$/h and box size $144$~Mpc/h. We chose an initial level of refinement $\ell=9$ ($512^3$), 
but we allowed for refinement down to a maximum level $\ell_{\rm max}=16$. Then, we identified dark matter halos with the AdaptaHOP algorithm 
\citep{2004MNRAS.352..376A}, using the version implemented and tested by \cite{2009A&A...506..647T}. We used the results from the halo finder to construct a set of 51 halos whose 
{\it total} masses are $M_{\rm tot}>10^{14}$~M$_\odot$ and whose neighboring halos do not have masses larger than $M/2$ within a spherical region of five times their virial radius. 
In our previous analysis \citep{2014MNRAS.440.2290M}, we determined that only 25 of these clusters are relaxed. 
High resolution initial conditions were extracted for each of the 51 halos, with the aim of performing zoom-in re-simulations. We ran three different re-simulations per halo: 
(I) including dark matter and neglecting baryons, (II) including dark matter and baryons and stellar feedback, (III) including baryons, stellar feedback and AGN feedback. In this paper we focus 
on cases (II) and (III), because they allow to study the properties of BCGs.

In the re-simulations, the dark matter particle mass is $m_{\rm cdm}=1.62\times 10^{8}$~M$_\odot$/h, while the baryon resolution element has a mass of $m_{\rm gas}=3.22\times 10^{7}$~M$_\odot$. 
We set the maximum refinement level to $\ell=17$, corresponding to a minimum cell size $\Delta x_{\rm min} = L/2^{\ell_{\rm max}}\simeq 1.07$ kpc/h. The grid was dynamically refined using a quasi-Lagrangian approach: 
when the dark matter or baryonic mass in a cell reaches 8 times the initial mass resolution, it is split into 8 children cells. Table~\ref{tab:mass_par} summarizes the particle mass and spatial resolution achieved in 
our simulations.

We model gas dynamics using a second-order unsplit Godunov scheme \citep{Teyssier:2002p451, Teyssier:2006p413, Fromang:2006p400} based on the HLLC Riemann solver and the
MinMod slope limiter \citep{Toro:1994p1151}. We assume a perfect gas equation of state (EOS) with polytropic index $\gamma=5/3$. All the zoom-in runs include sub-grid models for gas cooling which account for H, He and 
metals and that use the \citealt{sutherland_dopita93} cooling function. We directly follow star formation and supernovae feedback (``delayed cooling" scheme, \citealt{Stinson:2006p1402}) and metal enrichment. 
In the 51 re-simulations we also included an AGN feedback scheme, a modified version of the \cite{Booth:2009p501} model. In this scheme, supermassive black holes (SMBHs) are modeled as sink particles and AGN feedback is 
provided in form of thermal energy injected in a sphere surrounding each SMBH. We label these simulations as AGN-ON. The other 51 HYDRO simulations do not include AGN feedback. We label these simulations as AGN-OFF. 
More details about the AGN feedback scheme can be found in \cite{2011MNRAS.414..195T} and \cite{2012MNRAS.420.2859M}.

\begin{table}
\begin{center}
{\bfseries Mass and spatial resolution}
\begin{tabular}{|l|c|c|c|}
\hline
\hline
{\itshape Type} & $m_{\rm cdm}$&  $m_{\rm gas}$ & $\Delta x_{\rm min}$ \\
 & $[10^{8}$ M$_\odot$/h] & $[10^{7}$ M$_\odot$/h] & [kpc/h] \\
\hline
\hline
 Original box & $ 15.5 $ & n.a. & $2.14$ \\
 Zoom-in & $1.62$ & $3.22$ & $1.07$ \\
\hline
\hline
\end{tabular}
\end{center}
\caption{Mass resolution for dark matter particles, gas cells and star particles, and spatial resolution (in physical units) for our simulations. }\label{tab:mass_par}
\end{table}

There are a number of unconstrained free parameters in the galaxy formation model we adopt. In particular, the efficiency of the AGN feedback and star formation process are two crucial parameters. A careful study of the tuning 
of AGN feedback models implemented in the {\scshape ramses} code has been performed by \cite{2012MNRAS.420.2662D}. In our case, the tuning has been performed re-simulating one of the halos in our catalog (the less massive one) 
several times while varying the star formation efficiency $\epsilon_*$ and the size of the region where the AGN feedback energy is injected. The model that best reproduces the $M_{\rm BH}-\sigma$ relation and the 
central galaxy masses has star formation efficiency $\epsilon_*=0.03$ and size of the AGN feedback injection region equal to twice the cell size. For the AGN-OFF simulations we adopt $\epsilon_*=0.01$, which is close to the lower 
limit of the observed star formation efficiencies. However, even with such a low $\epsilon_*$, overcooling is still expected to be very strong in galaxy clusters, leading to BCG properties that disagree with the observations 
(see following Sections). 

In a previous study \citep{2012MNRAS.420.2859M}, we showed that AGN feedback plays a relevant role in the evolution of simulated BCGs. The limit of our 
previous work was that the analysis was limited to one cluster, simulated at very high resolution. The large sample we analyse in this paper constitutes a complementary data set and allows to test our galaxy formation model in a 
large number of halos with different merger histories.

\section{Properties of the simulated BCGs}

In this Section we analyse the properties of the BCGs in our sample. Given that approximately half of our simulated clusters are relaxed whereas the other half are unrelaxed, our BCGs 
should represent a fairly unbiased sample of objects at $z=0$. The resolution of our simulations is not adequate for the comparison of the progenitors of the BCGs to high redshift galaxies.  
On the other hand, at $z=0$ the BCGs are better resolved, therefore we focus our analysis on the $z=0$ properties.

\subsection{BCG identification and analysis}\label{sec:BCGdef}

BCGs are extremely massive systems sitting close to cluster centres. Since their position can be offset with respect to the cluster centre (see e.g. \cite{2014MNRAS.tmp..335M}), 
we identify BCG centres separately from halo centres. To accomplish that we ran a modified implementation of AdaptaHOP \citep{2004MNRAS.352..376A,2009A&A...506..647T} on the 
distribution of stellar particle with the aim of identifying galaxy centres. The BCG centres were identified as the centres of the most massive galaxies close to the cluster cores. 
We use the information provided by AdaptaHOP to remove satellite galaxies from our analysis and keep only the BCG surrounded by its extended stellar halo, i.e. 
the Intra-Cluster Light (ICL). 

The recent work by \cite{2013MNRAS.436.1750R} explores the properties of BCGs in smoothed particle hydrodynamics (SPH) simulations. These authors define BCG masses in several ways and 
reach qualitatively similar conclusion for all their mass definitions. Their results show that dynamical decoupling of 
BCG and ICL is a very accurate procedure to measure BCG stellar masses in simulations \citep{Puchwein:2010p763, 2014MNRAS.437..816C}, however, when the aim is measuring the integrated properties of 
BCGs, it is much easier to rely on mock photometric data. For this reason, we base our analysis on mock V-band images of the BCGs in our sample generated using the {\scshape sunset} code 
included in the {\scshape ramses} package. {\scshape sunset} is based on a simplified STARDUST SSP model \cite{1999A&A...350..381D}. 

Once the V-band images are generated we measure surface brightness profiles for all the BCGs. The BCG mass is defined as the projected mass enclosed within the isophotal contour with 
$\mu_{\rm V}=25$ mag/arcsec$^{-2}$. We ought to stress that this threshold is one mag/arcsec$^{-2}$ fainter than the one adopted by \cite{2013MNRAS.436.1750R}, however this allows us to 
include stellar mass that might be lost by adopting a lower (brighter) threshold. As a matter of fact, recent studies suggest \citep{2005ApJ...618..195G, 2014arXiv1401.7329K} show that the 
stellar light distribution of BCGs is much more extended than assumed in previous works (see the next Section for a discussion on this topic). The outer isophotal contour and the measured stellar mass are then used to estimate half-light 
radii, velocity dispersions and star formation rates (SFRs) for all the BCGs. To avoid biases related to the choice of a line-of-sight, we repeat the same analysis along 10 random 
lines-of-sight for each galaxy, then we average the results. 

\begin{figure*}
    \includegraphics[width=0.49\textwidth]{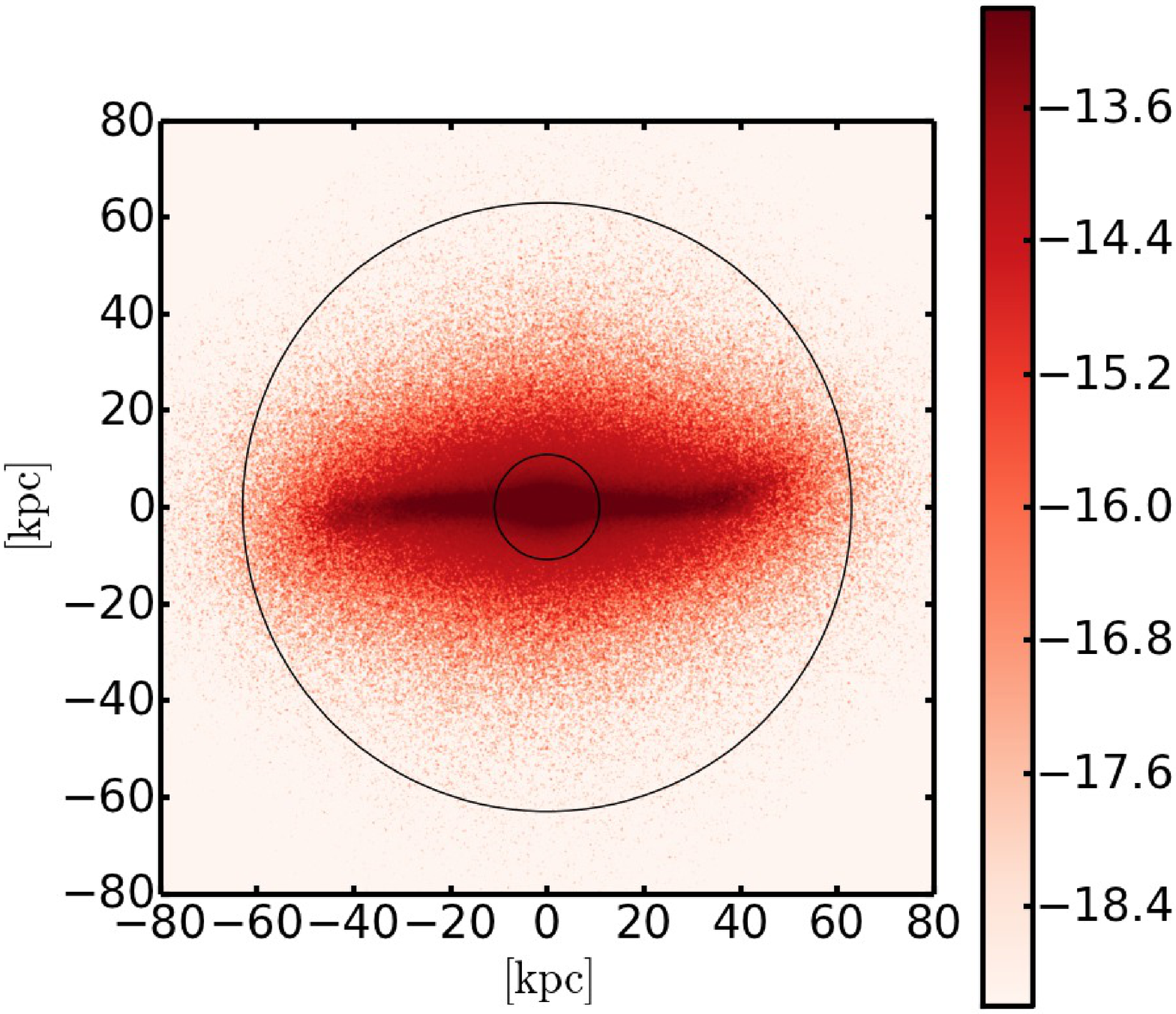}
    \includegraphics[width=0.49\textwidth]{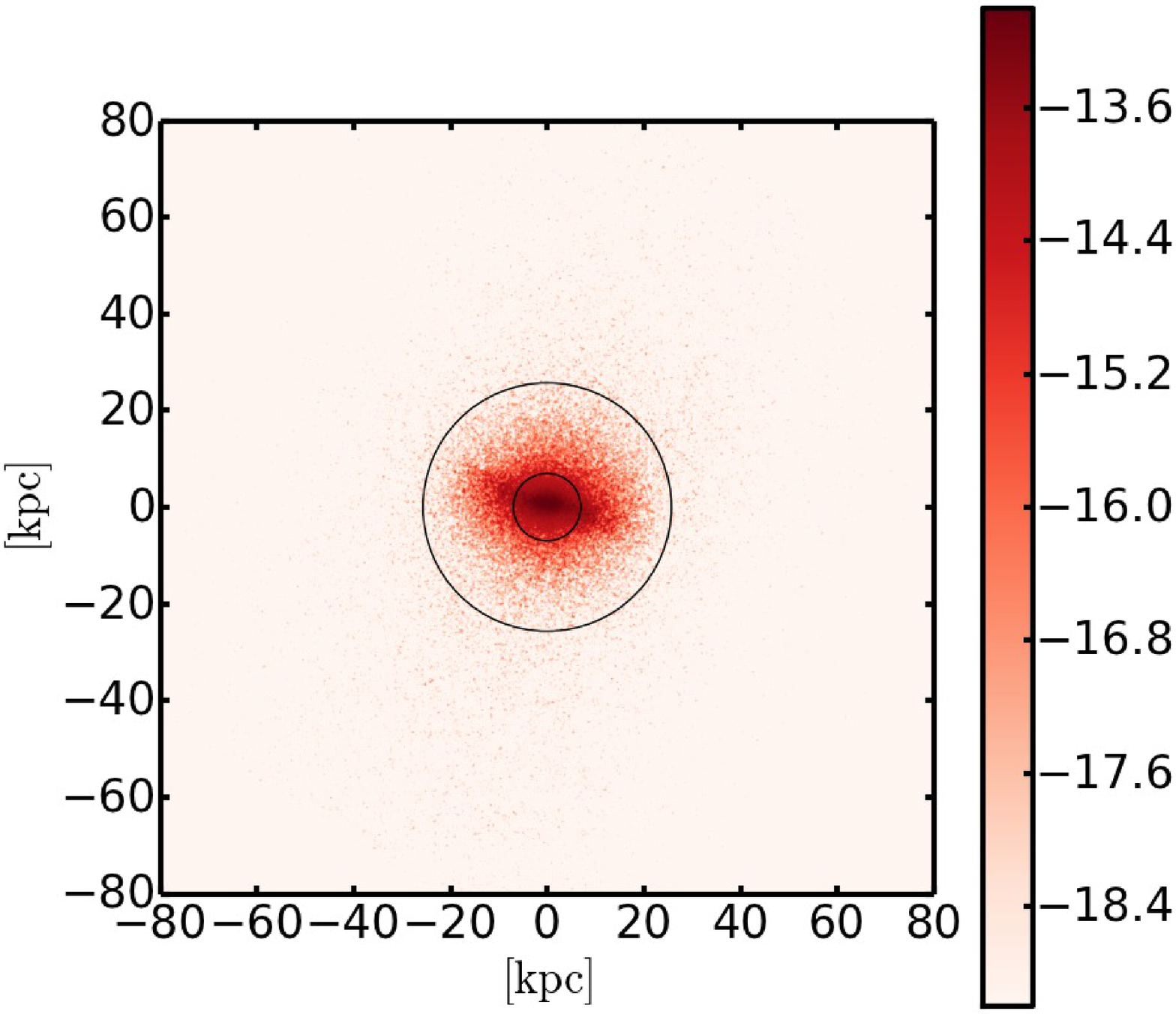}
\caption{Edge-on view of one of the BCGs in our sample: V-band flux in arbitrary units. Left: AGN-OFF. Right: AGN-ON. The colour scale is the same in both panels. All the satellites have been removed from these images.}
  \label{fig:images}
\end{figure*}

Figure~\ref{fig:images} shows an edge on image of one of the BCGs in our sample. To make these images we use the information from the halo finder to remove the contribution from satellite galaxies. 
The image on the left represents the AGN-OFF case, whereas the image on the right represents the AGN-ON case. 
Significant differences are noticeable in the luminosity and morphology of the object. The AGN-OFF BCG is much more extended and luminous than the AGN-ON realization in the same halo. 
Furthermore, the AGN-OFF BCG is a disk galaxy, whereas the AGN-ON BCG appears to be a very luminous elliptical galaxy.

In the following Sections, we will show direct quantitative comparison of the two scenarios to observational results. This comparison will help in gaining a better understanding of the 
the models and their predictions.

\subsection{Halo mass versus stellar mass}

Successful models of galaxy formation are expected to reproduce the relation between galaxy stellar mass and halo mass. The slope of this relation is not constant, 
it varies with halo mass \cite{2003MNRAS.340..771V, 2009ApJ...696..620C, 2009ApJ...699.1333H, Moster:2010p5423, 2013MNRAS.428.3121M, 2013ApJ...770...57B, 2014arXiv1401.7329K}. 
This fact can be interpreted as evidence that halos of different masses convert their initial baryonic content into stars at different rates. Traditional galaxy formation models are known 
to produce gas over-cooling in galaxy clusters. This process leads to very efficient conversion of gas into stars, i.e. to star formation rates and galaxy masses in excess with respect 
to those observed cluster galaxies. AGN feedback has been proposed has a mechanism to solve the over-cooling problem and the over-production of stars simultaneously 
\citep{Springel:2005p1553, croton_etal06, Sijacki:2007p1032, Booth:2009p501}. In this Section, we compare the results of our simulations to the most recent halo mass versus stellar 
mass relations reported in the literature.  

\begin{figure*}
    \includegraphics[width=0.49\textwidth]{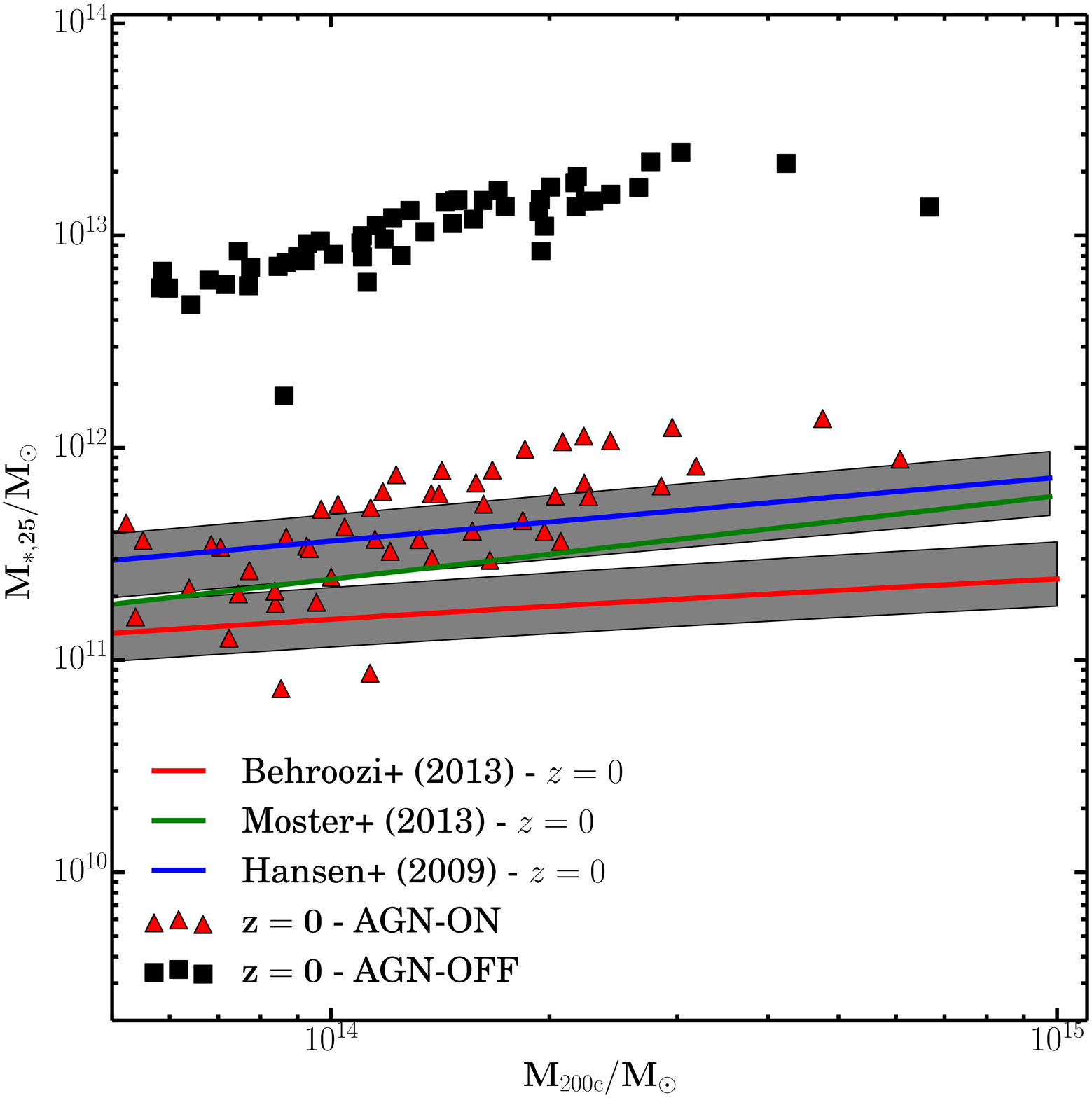}
    \includegraphics[width=0.49\textwidth]{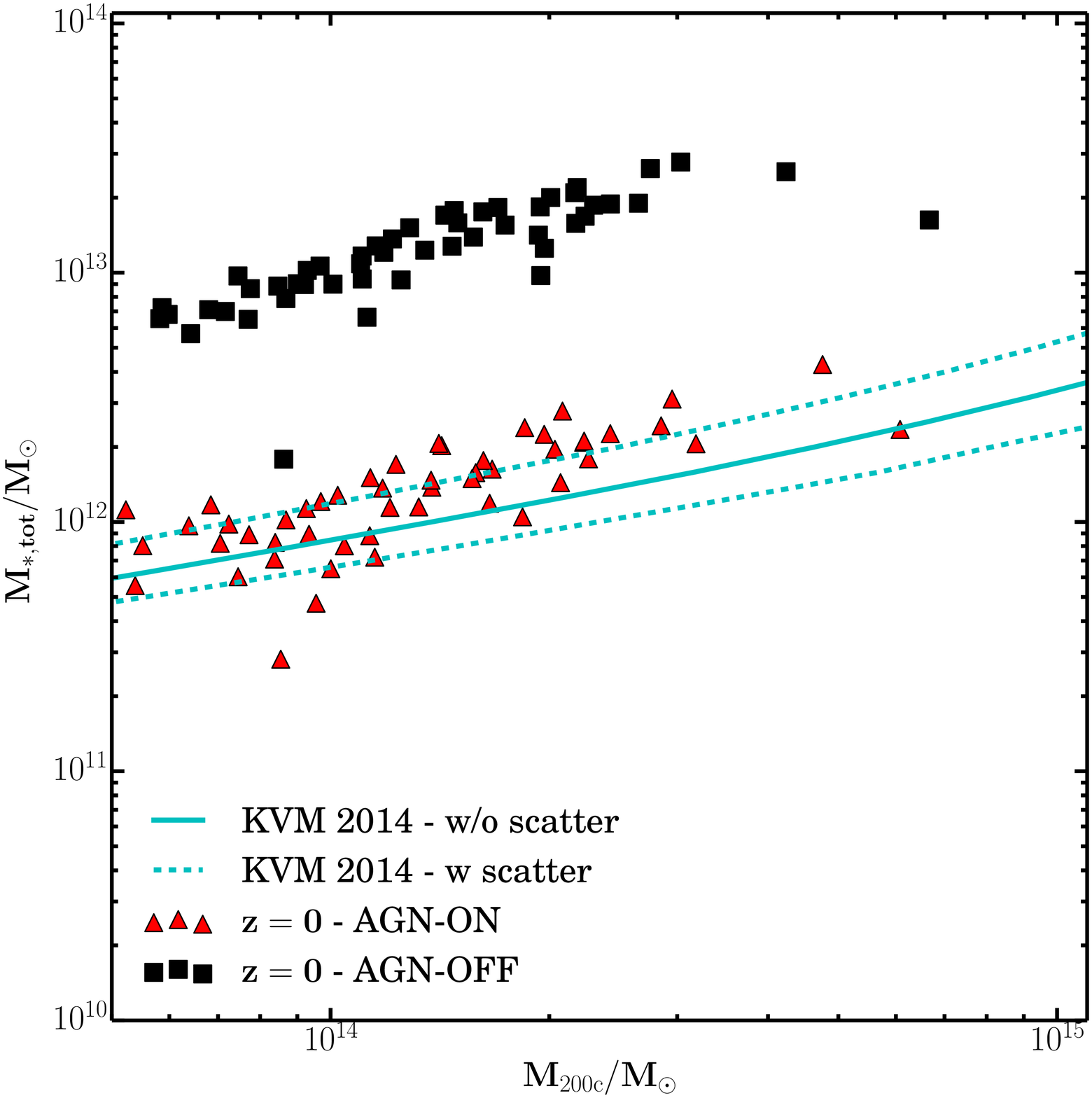}
\caption{Left: Results from our simulations versus the halo mass vs. stellar mass relation at redshift $z=0$ from Hansen et al. (2009) (blue line, with its $1\sigma$ scatter), 
Moster et al. (2013) (green line), Behroozi et al. (2013) (red line, with its $1\sigma$ scatter). Our AGN-ON simulations are represented 
by the red triangles, whereas the AGN-OFF simulations are represented by black squares. The BCG stellar mass definition assumes a surface brightness limit $\mu_{\rm V}=25$ mag/arcsec$^{-2}$. Right: 
Halo mass vs. stellar mass relation at redshift $z=0$ from Kravtsov et al. (2014) (cyan line) versus our simulations. Our AGN-ON simulations are represented 
by the red triangles, whereas the AGN-OFF simulations are represented by black squares. The stellar mass on the y axis is the total stellar mass associated to the BCG+ICL component.}
  \label{fig:abmatch}
\end{figure*}

The left panel of Figure~\ref{fig:abmatch} shows the comparison of our simulated BCGs at redshift $z=0$ to the halo mass versus stellar mass relation obtained using several techniques 
\citep{2009ApJ...699.1333H, Moster:2010p5423, 2013MNRAS.428.3121M, 2013ApJ...770...57B, 2014arXiv1401.7329K}. AGN-OFF simulations produce central galaxies that are $\sim 10$ times more 
massive than what is expected to observe in the real Universe. As already mentioned, this is one of the effects of gas over-cooling. The AGN-ON results are in much better agreement 
with the expected halo mass vs. stellar mass relation for these massive halos, however this result needs to be more carefully analysed. The AGN-ON BCGs tend to match better the relations 
that predict a higher halo mass (Moster et al. 2013, Hansen et al. 2009). In the low mass range showed in this plot, our AGN-ON also agree with the results of Behroozi et al. (2013). 
The differences between the curves from Hansen et al. (2009), Moster et al. (2013) and Behroozi et al. (2013) are caused by different systematic effects, however they all lie within 
$2\sigma$ from each other, where $\sigma$ is the scatter in the halo mass vs. stellar mass relation (grey shaded areas in the left panel of Figure~\ref{fig:abmatch}). Given these consideration and 
{\itshape given our mass definition} (see Subsection~\ref{sec:BCGdef}), we cannot tell whether the AGN-ON simulations over-produce or under-produce BCG stellar masses.

\begin{figure*}
    \includegraphics[width=0.49\textwidth]{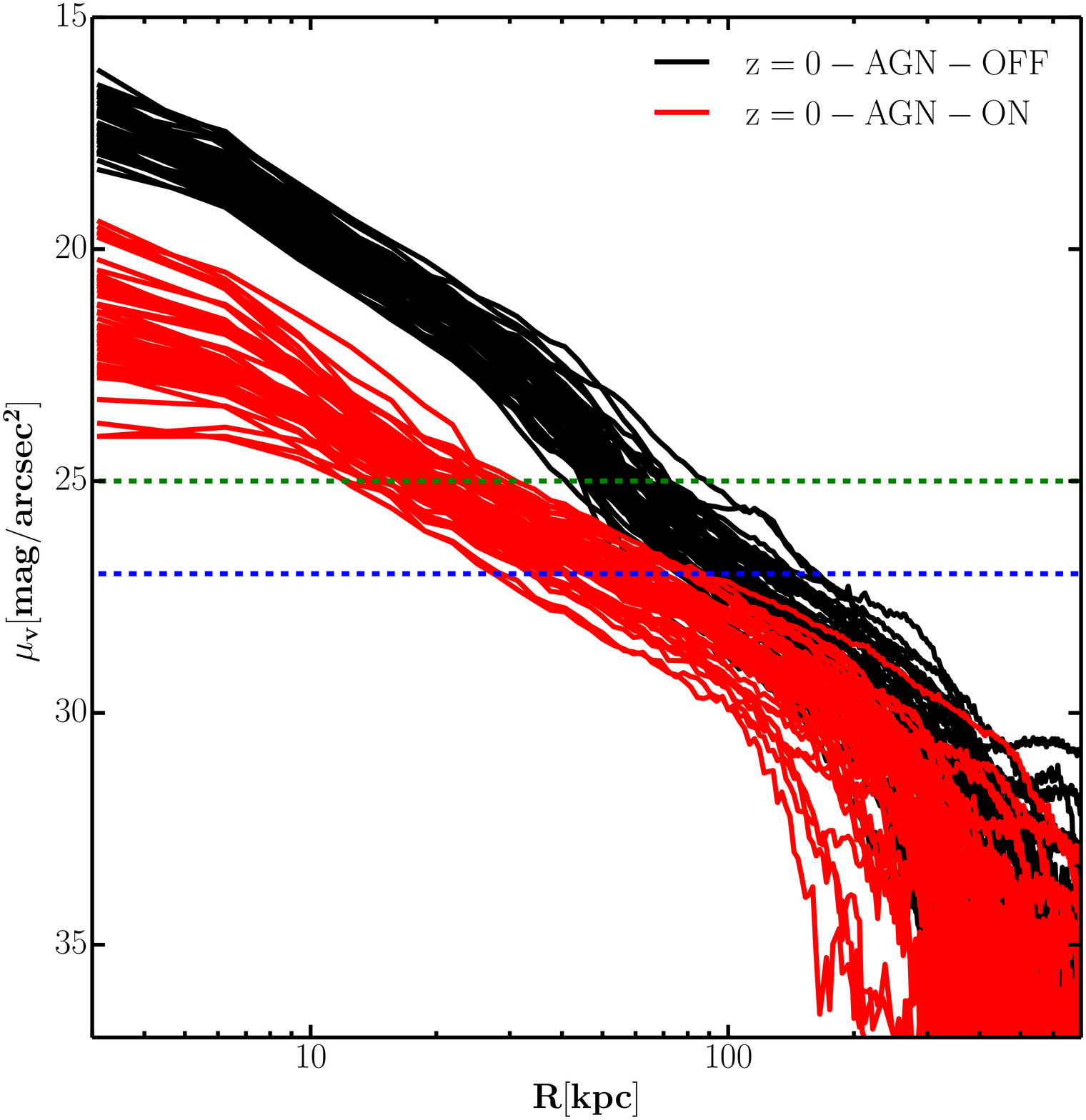}
    \includegraphics[width=0.49\textwidth]{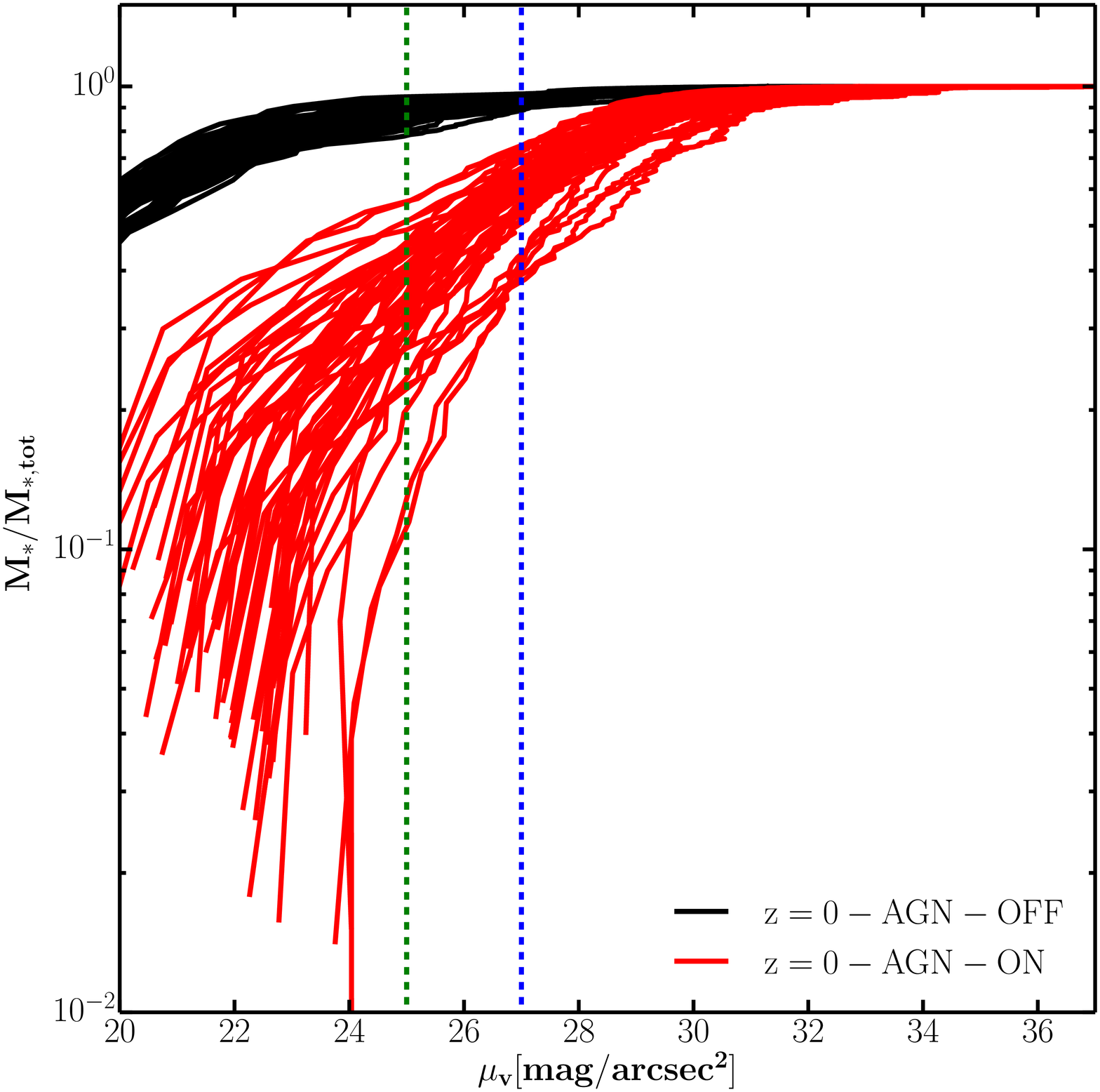}
\caption{Left: V-band surface brightness for all the BCGs in our sample. AGN-ON is shown in red, AGN-OFF is shown in black. The green dashed line represents the surface brightness cut adopted to 
define galaxy masses $\mu_{\rm V}=25$ mag/arcsec$^{-2}$. The blue dashed line represents an alternative cut $\mu_{\rm V}=27$ mag/arcsec$^{-2}$ which allows to include most of the 
Intracluster Light in the stellar mass estimate. Right: normalized stellar mass as a function of the adopted magnitude cut for all the BCGs. The mass distribution includes BCG+ICL. 
The green and blue lines represent $\mu_{\rm V}=25$ mag/arcsec$^{-2}$ and $\mu_{\rm V}=27$ mag/arcsec$^{-2}$, respectively.}
  \label{fig:sb}
\end{figure*}

Kravtsov et al. (2014) discuss in detail how abundance matching should be done in the mass range of BCGs: their halo mass vs. stellar mass has been obtained by using the recent 
calibration of the stellar mass function by \cite{2013MNRAS.436..697B}, which is based on accurate fits to the light profiles of the most luminous galaxies. For BCGs this means 
fitting the light profiles to distances up to $\sim 100 -200$ kpc from their centre. Such distance is $\sim 4-5$ times more extended than the one usually adopted to define BCG sizes, and 
is likely to include part of the Intracluster Light (ICL) in the central object mass budget. This leads to the increased stellar mass values observed in the stellar mass vs. halo mass 
relation that we also show in the right panel of Figure~\ref{fig:abmatch}. Therefore, the mass definition based on the $\mu_{\rm V}=25$ mag/arcsec$^{-2}$ threshold is not adequate for comparison to 
Kravtsov et al. (2014), because it underestimates the mass in the BCG+ICL component. With the aim of studying the effect of adopting a given surface brightness threshold, we carefully 
inspected the relationship between mass profiles and light profiles in our simulations.

\begin{figure}
    \includegraphics[width=0.49\textwidth]{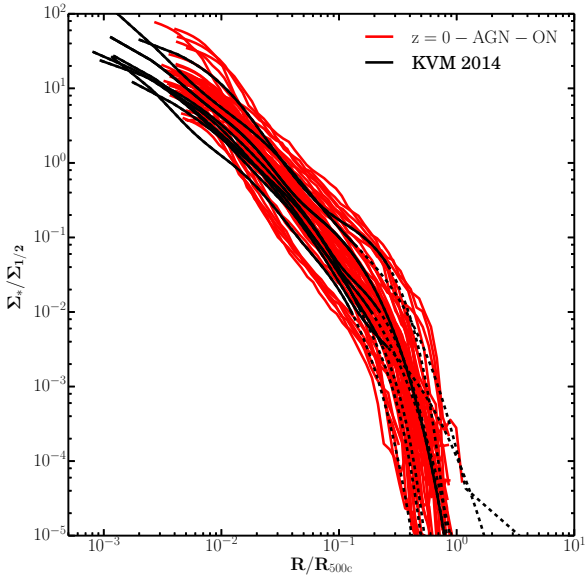}
\caption{Normalized surface density profiles of the AGN-ON BCGs (red lines) compared to those of the observed BCGs of Kravtsov et al. (2014). Solid black lines represent the fits in the region where the 
profiles have been measured, dashed black lines represent the fits in the regions where the profile has been extrapolated. }
  \label{fig:norm_prof}
\end{figure}

The left panel of Figure~\ref{fig:sb} shows the surface brightness profiles of the AGN-ON and AGN-OFF BCG. Most of the light is concentrated within the $\mu_{\rm V}=25$ mag/arcsec$^{-2}$ limit, however there 
is still a significant amount of light outside of this region out to $\sim 60-100$ kpc. To measure the amount of mass that is missed by an observation with a given surface brightness limit, 
we plot the enclosed stellar mass normalized to the total stellar mass in BCG+ICL as a function of the surface brightness limit adopted in the right panel of Figure~\ref{fig:sb}. 
The total stellar mass in BCG+ICL is obtained by removing the stellar mass in the satellite galaxies using information from the halo finder. The AGN-OFF 
galaxies are much more centrally concentrated, so most of the mass ($\sim 80-90$ \%) is still enclosed in the isophotes with $\mu_{\rm V}<25$ mag/arcsec$^{-2}$. In the AGN-ON case the stellar 
mass distribution is less centrally concentrated and the BCGs have a very extended ICL. As a result, very high $\mu_{\rm V}$ values are needed to be able to observe the ICL mass. A limit 
$\mu_{\rm V}=25$ mag/arcsec$^{-2}$ might be adequate to capture the mass of the BCG, but not the entire mass of the BCG+ICL. Our results show that $\mu_{\rm V}=25$ mag/arcsec$^{-2}$ 
captures only $\sim 10-60$ \% of the total stellar mass. A surface brightness limit $\mu_{\rm V}=27$ mag/arcsec$^{-2}$ gives $40-80$ \% of the total stellar mass in BCG+ICL. If we use the 
{\itshape total} stellar mass in BCG+ICL and compare to Kravtsov et al. (2014) we get the result in the right panel of Figure~\ref{fig:abmatch}: the match of our simulations to these results is excellent. 
This result is non-trivial since the Kravtsov et al. (2014) masses are obtained by extrapolating their multiple Sersic fits to high radii. Figure~\ref{fig:norm_prof} shows normalized surface density profiles 
for the AGN-ON BCGs compared to those of the observed BCGs of Kravtsov et al. (2014). Radius has been normalized with respect to $R_{\rm 500c}$, surface density has been normalized with respect to 
the half-mass surface density $\Sigma_{1/2}=0.5M_{\rm *,tot}/4\pi R_{1/2}^2$, where $R_{1/2}$ is the BCG+ICL half-mass radius and $M_{\rm *,tot}$ is the stellar mass in BCG+ICL. Figure~\ref{fig:norm_prof} 
shows that the extrapolation of the profiles in the observational sample closely matches the profiles of the simulated BCGs at high radii. This implies that our simulations closely match observations and that 
extrapolation of fitted surface density profiles (triple Sersic functions in this case) are sufficiently robust to be used to measure stellar masses.

%%%%%%%%%%%%%%%%%%%%%%%%%%%%%%%%%%%%%%%%%%%%%%%%

\section{Other Properties of the simulated BCGs}

\begin{figure*}
    \includegraphics[width=0.49\textwidth]{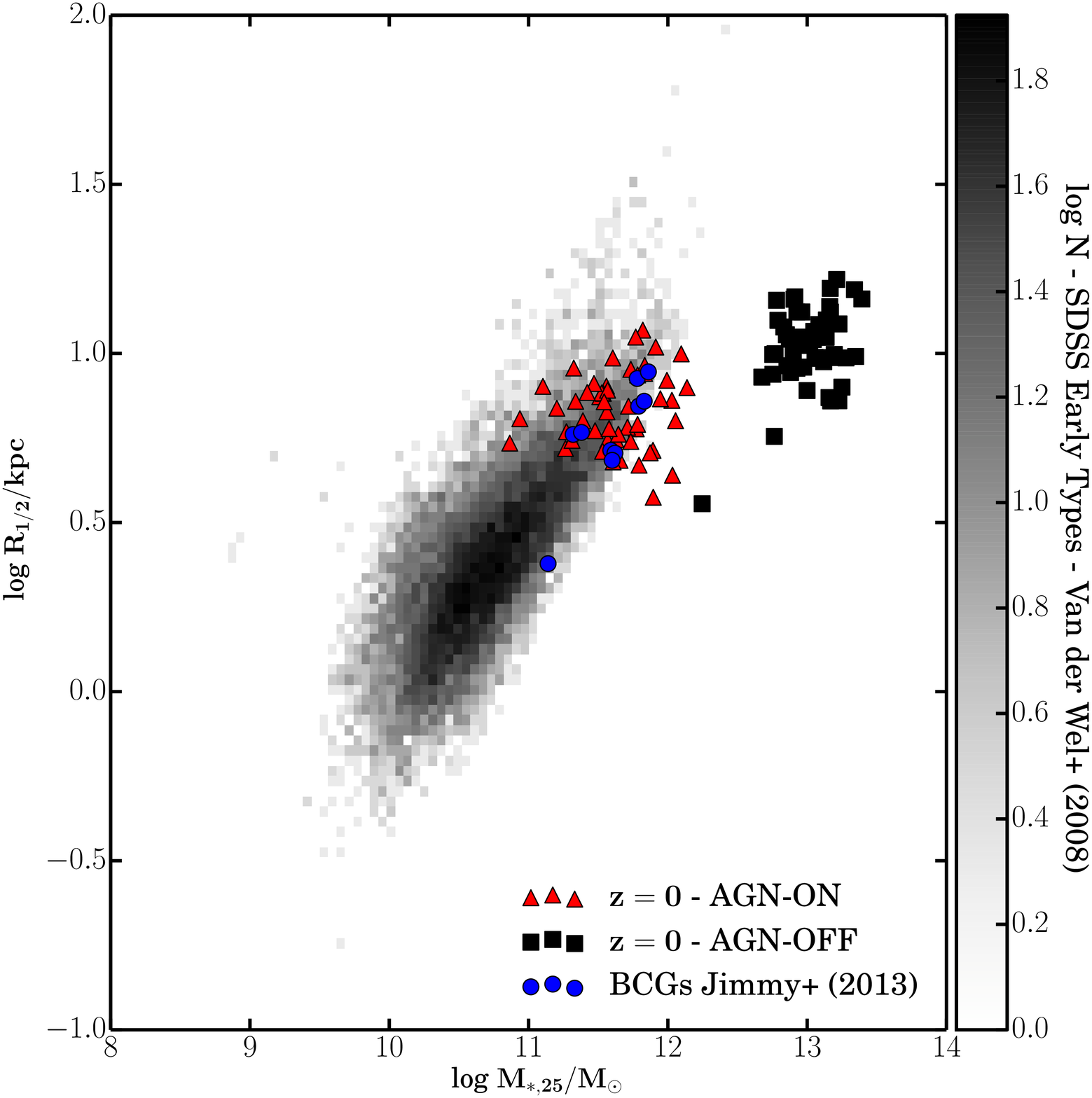}
    \includegraphics[width=0.49\textwidth]{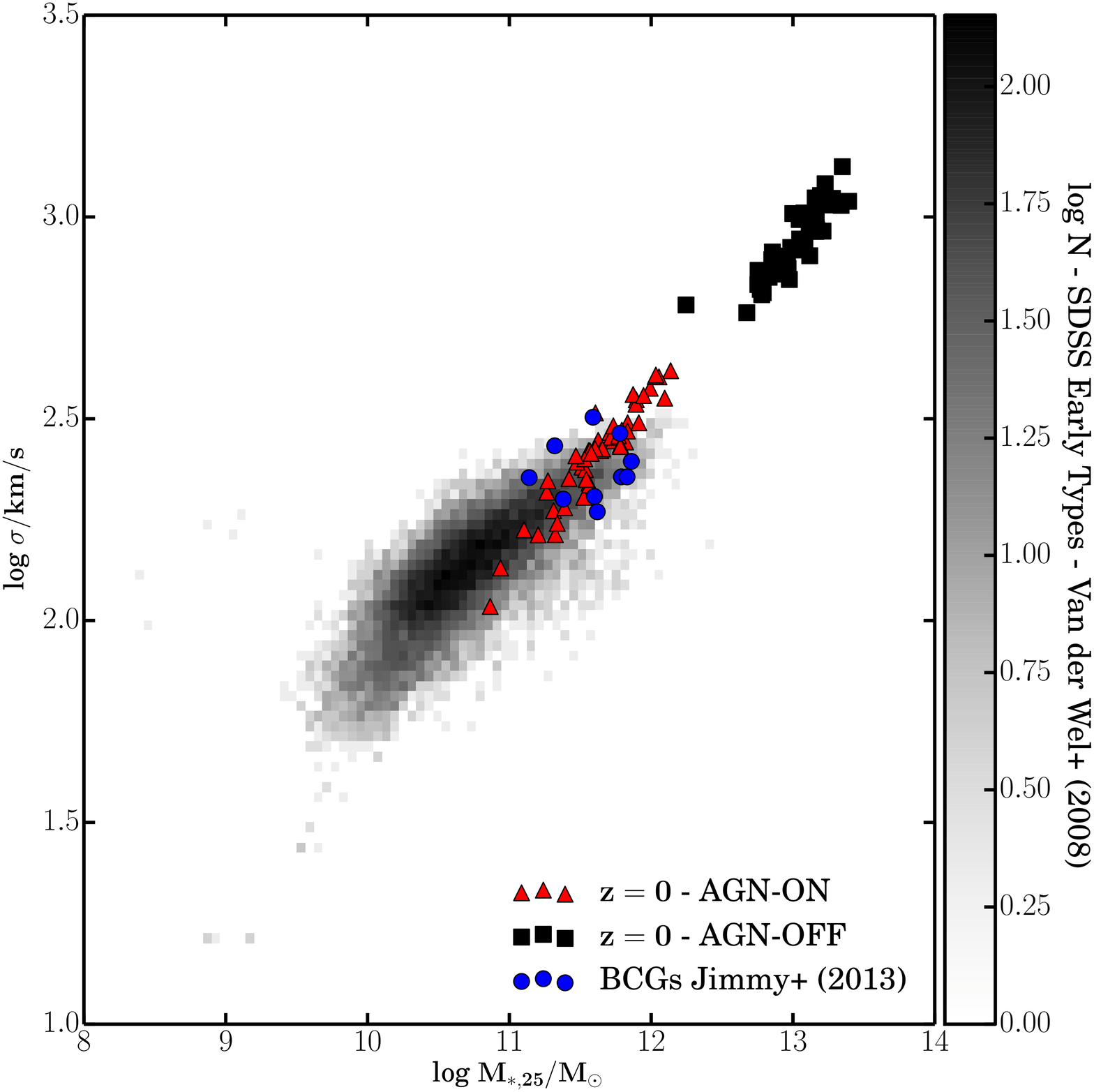}
\caption{Left: Stellar mass vs. size. The size is the half-light radius for the simulated BCGs and the effective radius for the observational data. Right: Stellar mass vs. stellar velocity dispersion. 
The shaded area represents galaxy counts per bin from the ETG sample by Van der Wel et al. (2008) sample from SDSS. The blue circles are the BCGs analysed by Jimmy et al. (2013).}
  \label{fig:mass_sigma_reff}
\end{figure*}

In this Section we compare the properties of the simulated BCGs obtained by integrating the information within the half-light radius. We define the half-light radius as the radius containing half of the luminosity 
of the BCG within the isophote with $\mu_{\rm V}<25$ mag/arcsec$^{-2}$. The stellar mass of the BCG has been measured by using the criterion described in the previous section. Star formation rates (SFR) have been 
measured considering only the stars formed in the latest $5\times 10^8$ yr. Half-light radii, velocity dispersions, stellar masses are compared to the values found 
in observations. We measure these parameters by averaging over several lines of sight per galaxy. 
The goal of such comparison is to test the galaxy formation prescriptions and to identify discrepancies with the observations that point at evident flaws in the model.

The left panel of Figure~\ref{fig:mass_sigma_reff} shows the stellar mass vs. size relation at $z=0$ for the simulated BCG compared to that found by \cite{vanderWel:2008p5391} for a sample of early-type galaxies 
extracted from the Sloan Digital Sky Survey (SDSS) and to the recent data published by \cite{2013ApJ...778..171J}. A similar comparison is shown in the right panel of the same figure but for the stellar mass vs. 
velocity dispersion relation. As we already know from the analysis in the previous Section, the masses of the AGN-OFF BCGs are one order of magnitude larger than the ones of massive galaxies observed in the real 
Universe. Figure~\ref{fig:mass_sigma_reff} also shows that the AGN-OFF galaxies are also larger, have much higher velocity dispersion and are more centrally concentrated than the AGN-ON galaxies. 
Gas over-cooling leads very efficient gas fueling of the BCG and to high star formation rates in disk structures (see left panel of Fig~\ref{fig:images}); due to the large amount of baryonic matter in their central 
regions, these BCGs become more centrally concentraded and obtain huge velocity dispersions. On the other hand the AGN-ON BCGs show a remarkable agreement in mass, size and velocity dispersion compared to 
the most massive early-type galaxies in \cite{vanderWel:2008p5391} and to the BCGs in \cite{2013ApJ...778..171J}. A small discrepancy can be noticed between the sizes of the most massive BCGs and those of the 
early-type galaxies which could point at some flaw in the model that manifests its effects only in the most massive halos. However, we cannot draw a definitive conclusion on the actual existence of this discrepancy since it is 
known that the most massive BCGs are not scaled-up versions of elliptical galaxies \citep{2007MNRAS.379..867V}. 

\begin{figure}
    \includegraphics[width=0.49\textwidth]{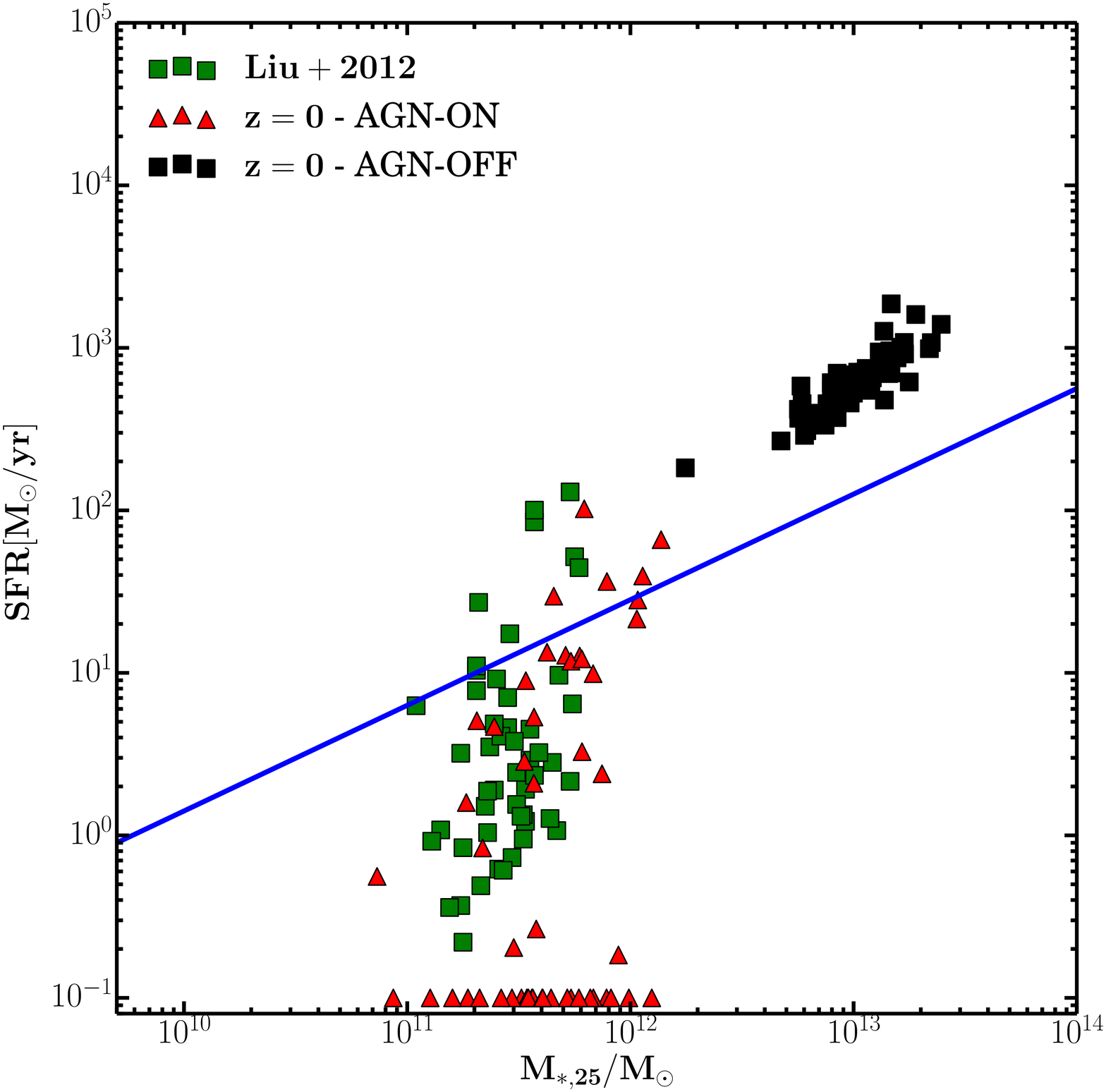}
\caption{Stellar mass vs. star formation rate for the \textquotedblleft star-forming\textquotedblright BCGs. The simulations are compared to the observational data by Liu et al. (2012) (green squares). 
  The BCGs with reported SFR$<10^{-1}$ M$_{\odot}/yr$ are represented by upper limits (the measured SFR is 0 M$_{\odot}/yr$). The blue solid line represents a power-law extrapolation of the sequence for star forming 
  galaxies measured by Brinchmann et al. (2004).}
  \label{fig:mass_sfr}
\end{figure}

Interesting results are found when we compare the stellar mass vs. SFR relation at $z=0$ for the simulated BCGs to the data of \cite{2012MNRAS.423..422L} (SFRs of early-type BCGs measured from their H$\alpha$ emission, 
SDSS data), as shown in Figure~\ref{fig:mass_sfr}. AGN-OFF galaxies have SFRs too high by a factor $5-10$ with respect to the most intensely star-forming BCGs in the \cite{2012MNRAS.423..422L} sample. The AGN-ON BCGs 
show an interesting dicothomy: most of these galaxies have extremely low SFRs shown as upper limits in the figure (SFR$<10^{-1}$ M$_{\odot}/yr$), i.e. they are completely quenched BCGs; a secondary population of 
\textquotedblleft star-forming\textquotedblright BCGs is also observed, in remarkable agreement with the data of \cite{2012MNRAS.423..422L}. The BCGs with the highest star-formation rates are close to the sequence of 
star-forming galaxies at redshift $z=0$ (the blue line in Figure~\ref{fig:mass_sfr} is an extrapolation of the \cite{2004MNRAS.351.1151B} results in the shown mass range). The prescription adopted for the AGN scheme 
allows for late mild star formation events in BCGs, typically in the cluster with the most un-relaxed central regions. This star forming activity is likely to be a transient and to be suppressed by AGN feedback bursts 
of activity, since the gas accretion required for star formation also feeds the central supermassive black hole and will trigger an AGN burst. The dichotomy between mildly star-forming and quenched BCGs is observed 
also in the most massive BCGs which sit in the most massive halos. {Despite this qualitative match with observations, the fraction of star-forming BCGs in the simulated sample is $\sim 50 \%$, higher than the fraction measured by 
Liu et al. (2012) for an X-ray selected sample ($\sim 20 \%$). This discrepancy between observations and simulations and might be interpreted as a limit of the sub-grid model we adopt for AGN feedback. }

\begin{figure}
    \includegraphics[width=0.49\textwidth]{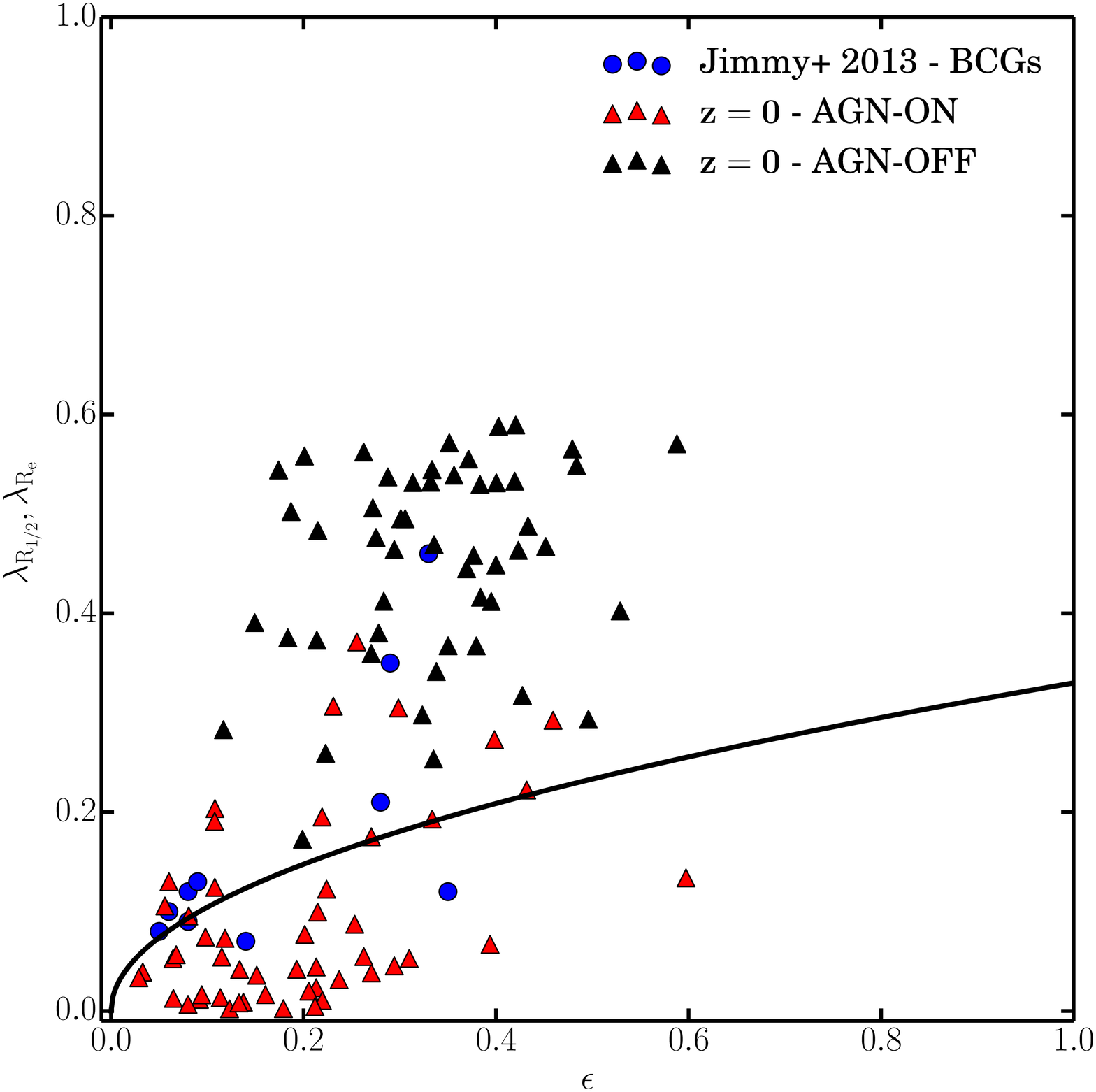}
\caption{Ellipticity (edge-on view) vs. angular momentum probe parameter $\lambda_{\rm R}$ measured within the half-light radius. This plot is used to separate fast rotators and slow rotators: here the separation is represented by the black solid line 
$\lambda_{\rm R}=0.33*\sqrt{\epsilon}$. Values measured within the effective radius of a sample of real BCGs from Jimmy et al. (2013) are also compared to our results (blue circles). The Jimmy et al. ellipticities are apparent and have 
not been corrected. }
    \label{fig:kinematics}
\end{figure}

Finally, we compare the kinematic properties of the simulated objects to those of observed BCGs at redshift $z=0$. One of the most interesting plots for early-type galaxies is the $\epsilon-\lambda_{\rm R}$ 
\citep{Emsellem:2007p5467, 2011MNRAS.414..888E}, typically studied using integral field spectroscopy: $\epsilon $ is the galaxy ellipticity, $\lambda_{\rm R}$ is a parameter that quantifies the angular momentum 
content in galaxies, it has been specifically defined to be measured from integral field spectroscopy and it is defined as 
\begin{equation}
\lambda_{\rm R}=\frac{\langle R |V| \rangle}{\langle R \sqrt{V^2+\sigma^2} \rangle},
\end{equation}
where $V$ is the line of sight velocity in a pixel, $\sigma$ is the velocity dispersion in the pixel, $R$ is the distance of the pixel from the centre and the brackets represent an average over all the pixels. 
Low values of $\lambda_{\rm R}$ are associated to slowly rotating galaxies (slow rotators), high values of $\lambda_{\rm R}$ are associated to fast rotating galaxies (fast rotators). We 
measure the value of $\lambda_{\rm R}$ within the half-light radius and the edge-on ellipticities of the simulated BCGs, then compare our results to the BCGs observed by \cite{2013ApJ...778..171J}. {We ought to stress that \cite{2013ApJ...778..171J} report apparent ellipticities and that they measure $\lambda_{\rm R}$ within the effective radius. Despite these differences between the measurement method, observational uncertainties are still large enough  
to allow a meaningful comparison of data and simulations. }
The comparison of the simulated BCGs to 
the \cite{2013ApJ...778..171J} data is shown in Figure~\ref{fig:kinematics}. The black solid line $\lambda_{\rm R}\propto \sqrt{\epsilon}$ represents the separation between fast and slow rotators (Emsellem et al. 2011). The AGN-OFF 
population is exclusively composed of fast rotators in contrast with the properties of real BCGs. These AGN-OFF BCGs possess very massive rotating disk components. On the other end the AGN-ON BCGs show a variety of 
kinematic properties that is observed also in the \cite{2013ApJ...778..171J}: both fast and slow rotators are produced. {If we assume Poisson error bars, the fast-to-slow rotators ratio in the AGN-ON sample is $\sim 0.29$, whereas it is 
$\sim 33 \%$ in the \cite{2013ApJ...778..171J} sample. The fast-to-slow rotators ratio of AGN-ON BCGs is consistent with the observed one within the observational uncertainty. However, some considerations are needed before drawing conclusions.} First of all, the size 
of the \cite{2013ApJ...778..171J} sample is quite small and comparison to additional data samples need to be performed  (e.g. GAMA \citep{2014MNRAS.440..762O} or future surveys like MASSIVE). Second, the kinematic structure of the less 
massive BCGs in our simulations could be better resolved in higher resolution simulations like the one analysed in \cite{2012MNRAS.420.2859M}. Unfortunately, running one of these simulations requires roughly the same 
amount of computational resources as the entire sample analysed here. Third, the strategy we adopted to select the halos for the zoom-in simulations might introduce a bias in the selection of fast vs. slow rotators. 
We will carefully explore these issue in future work. 

\subsection{Comparison to previous literature}
 
{The formation of BCGs has been studied from a theoretical viewpoint by several authors which identified some of the main elements that drive the formation of these extremely massive galaxies. \cite{1998ApJ...502..141D} explored the problem of forming a very massive galaxy at the centre of a cluster using N-body simulations finding that mergers of several massive galaxies can lead to the formation of a large BCG. More recently, \cite{2007ApJ...668..826C} studied the build-up of BCGs and ICL and the evolution of the galaxy mass function between redshift $z=1$ and $z=0$, concluding that a large fraction ($\sim 80 \%$) of the stars from disrupted halos in a cluster end up contributing to the ICL. De \cite{DeLucia:2007p1656} studied the evolution of BCGs using semi-analytical modeling and concluded that most of their mass is assembled at redshift $z>1$ by rapid cooling flows which are later suppressed by AGN feedback. In our previous paper \citep{2012MNRAS.420.2859M}, we analysed the properties of the BCG formed at the centre of a $10^{14}$ M$_{\odot}$ cluster with a better resolution than the one we achieve in this paper (minimum cell size $\Delta x \sim 0.5$ kpc, dark matter mass particle $\sim 8\times 10^6$ M$_{\odot}$). Our previous results are basically consistent with those we show in this paper.

An interesting comparison can be made with the results of \cite{2013MNRAS.436.1750R} which performed an analysis very similar to the one done in this paper, but used SPH-based cosmological simulations. The resolution achieved in this work (softening length $2.5$ kpx/h, dark matter particle mass $\sim 8\times 10^8$ M$_{\odot}$) is slightly worse, but comparable to the one we achieve in this paper. Their implementation of AGN feedback is very similar to the one we adopt. Their conclusion is that AGN feedback is too efficient in the central regions of galaxies, generating density profiles with cores at their centre (see also \cite{2012MNRAS.422.3081M} and \cite{2012MNRAS.423.3243R}), but generally inefficient at reducing the global star formation in the most massive galaxies. The flattening of the density profiles is observed in both SPH and AMR simulations and is likely a feature produced by implementations of AGN feedback in form of thermal energy. In simulations, such flattening is measured at the scale of 5-10 kpc, but it is quite rare to observe flat density profiles in real elliptical galaxies at a scale larger than 3 kpc \cite{2012ApJ...756..159P}. This is an indication that this kind of implementation of AGN feedback is indeed too efficient at small scales. However, we do not find strong discrepancies in BCG masses as reported in \cite{2013MNRAS.436.1750R}. This difference might be related to intrinsic differences between SPH and AMR codes in the way gas of different phases mixes. This fact might influence the way gas cools, forms stars and accretes on SMBHs. Additionally, it is not guaranteed that implementations of the same sub-grid model work in the same way in an SPH and in an AMR code. These issues could be explicitly addressed by performing a dedicated code comparison project.}
 
%%%%%%%%%%%%%%%%%%%%%%%%%%%%%%%%%%%%%%%%%%%%%%%%%%%%%%%%%%%

\section{Summary and Conclusions}
\label{sec:summary}

In this paper we have analysed the low redshift properties of a sample of BCGs formed at the centre of 51 galaxy clusters simulated with the AMR code {\scshape ramses}. The zoom-in technique is adopted to obtain the 
required resolution only in the region surrounding each cluster, allowing us to save computing time. Several properties of the BCGs have been analysed, with the aim of testing the galaxy formation models adopted for 
the simulations. A similar analysis on AMR simulations has been shown in \cite{2012MNRAS.420.2859M}, however here we focus for the first time on much larger sample of AMR simulations of cluster evolution and BCG 
formation. One of the crucial effects needed to reproduce the properties that match those of real BCGs is a source of heating that slows down the cooling of large quantities of gas in massive galaxies, therefore 
quenching star formation. In this paper we specifically focus on the effect of AGN feedback as the heating source. The feedback scheme is very simple (it is largely inspired by \cite{Booth:2009p501}), so that 
careful comparison to observation can lead to the identification of its limits.

The results of this analysis can be summarized in a few points:
\begin{itemize}
  \item Simulations without AGN feedback do not reproduce the properties of observed BCGs. The simulated objects appear to be too massive, too large, too centrally concentrated and they are all fast rotators.
  \item Including AGN feedback reduces stellar masses and velocity dispersions. Galaxy half-light radii are only weakly modified. The properties of this population of simulated BCGs are very similar to those of real 
  BCGs observed at redshift $z=0$. There are some small discrepancies only for the most massive BCGs.
  \item The BCGs simulated in presence of AGN feedback are surrounded by a very extended ICL (up to a few $\sim 100$ kpc from the centre) that accounts for $20-60$ \% of stellar mass associated to the total 
  BCG+ICL component. A significant fraction of the ICL can be detected only by deep observations which probe regions of surface brightness $\mu_{\rm V}>27$ mag/arcsec$^{-2}$. However to detect the whole component even 
  deeper observations are needed.
  \item The objects in our sample match well the halo mass vs. stellar mass relation even when the comparison is performed by accounting for the total mass in BCG+ICL as in Kravtsov et al. (2014).
  \item {Approximately half of the BCGs simulated in presence of AGN feedback are completely quenched early-type objects. However, some of the BCGs form stars at rates similar to those in observed star-forming BCGs (Liu et al. 2012). The fraction of star-forming BCGs is a factor $\sim$2 lower in X-ray selected observational samples. This fact might be considered as an effect of partially inefficient quenching of star formation from the sub-grid model we adopt for AGN feedback.}
  \item The BCGs simulated in presence of AGN feedback are divided between fast and slow rotators as in the real Universe \citep{2011MNRAS.414L..80B, 2013ApJ...778..171J}. {The comparison made in this paper shows that 
  the fast-to-slow rotator fraction in the simulated sample is consistent with that measured by \citep{2013ApJ...778..171J}, however this result needs to be updated by considering large observational data samples and 
  simulations with improved resolution.}
\end{itemize}

The simple prescription for AGN feedback we adopted manages to solve the most critical problem for the formation of BCGs: gas over-cooling triggering excessive star formation at $z<1$. Given its simplicity (spherical 
symmetry of AGN activity, simple accretion rate formulae for supermassive black holes, lack of kinetic energy injection or jets, etc.) the model produces BCGs that closely match those in the real Universe. {The 
match appears to be somewhat worse when only the most massive BCGs are considered and the ratio of star-forming to non-star-forming BCGs is too high.} At the moment we cannot assess whether these discrepancies at the high mass end are caused by missing physical 
processes that might be relevant in galaxy clusters (anisotropic thermal conduction, jets, cosmic ray streaming). As time is needed before having larger observational samples of BCGs to carry out extensive comparison projects, 
there are at least two approaches to continue testing this particular galaxy formation framework. First, a fair sample of simulations with improved resolution might be needed to study the kinematics and formation 
histories of the most massive BCGs in the range of masses in which the largest discrepancies are observed. Second, a detailed analysis of the gas properties and accretion histories might be relevant because it will 
allow to quantify the exact condition in which over-cooling is still a problem and to directly compare to e.g. X-ray data. Given its relevance and need for careful discussion, we plan to perform this separate analysis 
in future work. 

\section*{Acknowledgments}
The simulations have been run on the Monte Rosa cluster at CSCS, Switzerland. Davide Martizzi is working as a postdoctoral fellow at University of California at Berkeley and his work is funded by the 
Swiss National Science Foundation. We thank Andrey Kravtsov for the useful discussions that improved the quality of this paper. We also thank our anonymous referee for his comments which helped us clarifying some critical points in the paper. 

%%%%%%%%%%%%%%%%%%%%%%%%%%%%%%%%%%%%%%%%%%%%%%%%%%%%%%%%%%%

\bibliography{paper_1.0}

\begin{thebibliography}{}

\bibitem[\protect\citeauthoryear{{Aubert}, {Pichon} \& {Colombi}}{{Aubert}
  et~al.}{2004}]{2004MNRAS.352..376A}
{Aubert} D.,  {Pichon} C.,    {Colombi} S.,  2004, \mnras, 352, 376

\bibitem[\protect\citeauthoryear{{Behroozi}, {Wechsler} \& {Conroy}}{{Behroozi}
  et~al.}{2013}]{2013ApJ...770...57B}
{Behroozi} P.~S.,  {Wechsler} R.~H.,    {Conroy} C.,  2013, \apj, 770, 57

\bibitem[\protect\citeauthoryear{{Bernardi}, {Meert}, {Sheth}, {Vikram},
  {Huertas-Company}, {Mei} \& {Shankar}}{{Bernardi}
  et~al.}{2013}]{2013MNRAS.436..697B}
{Bernardi} M.,  {Meert} A.,  {Sheth} R.~K.,  {Vikram} V.,  {Huertas-Company}
  M.,  {Mei} S.,    {Shankar} F.,  2013, \mnras, 436, 697

\bibitem[\protect\citeauthoryear{Bertschinger}{Bertschinger}{2001}]{Bertschinger:2001p1123}
Bertschinger E.,  2001, The Astrophysical Journal Supplement Series, 137, 1

\bibitem[\protect\citeauthoryear{Booth \& Schaye}{Booth \&
  Schaye}{2009}]{Booth:2009p501}
Booth C.~M.,  Schaye J.,  2009, Monthly Notices of the Royal Astronomical
  Society, 398, 53

\bibitem[\protect\citeauthoryear{{Brinchmann}, {Charlot}, {White}, {Tremonti},
  {Kauffmann}, {Heckman} \& {Brinkmann}}{{Brinchmann}
  et~al.}{2004}]{2004MNRAS.351.1151B}
{Brinchmann} J.,  {Charlot} S.,  {White} S.~D.~M.,  {Tremonti} C.,  {Kauffmann}
  G.,  {Heckman} T.,    {Brinkmann} J.,  2004, \mnras, 351, 1151

\bibitem[\protect\citeauthoryear{{Brough}, {Tran}, {Sharp}, {von der Linden} \&
  {Couch}}{{Brough} et~al.}{2011}]{2011MNRAS.414L..80B}
{Brough} S.,  {Tran} K.-V.,  {Sharp} R.~G.,  {von der Linden} A.,    {Couch}
  W.~J.,  2011, \mnras, 414, L80

\bibitem[\protect\citeauthoryear{{Conroy} \& {Wechsler}}{{Conroy} \&
  {Wechsler}}{2009}]{2009ApJ...696..620C}
{Conroy} C.,  {Wechsler} R.~H.,  2009, \apj, 696, 620

\bibitem[\protect\citeauthoryear{{Conroy}, {Wechsler} \& {Kravtsov}}{{Conroy}
  et~al.}{2007}]{2007ApJ...668..826C}
{Conroy} C.,  {Wechsler} R.~H.,    {Kravtsov} A.~V.,  2007, \apj, 668, 826

\bibitem[\protect\citeauthoryear{Croton, Springel, White, Lucia, Frenk, Gao,
  Jenkins, Kauffmann, Navarro \& Yoshida}{Croton et~al.}{2006}]{croton_etal06}
Croton D.~J.,  Springel V.,  White S.~D.~M.,  Lucia G.~D.,  Frenk C.~S.,  Gao
  L.,  Jenkins A.,  Kauffmann G.,  Navarro J.~F.,    Yoshida N.,  2006, \mnras,
  365, 11

\bibitem[\protect\citeauthoryear{{Cui}, {Murante}, {Monaco}, {Borgani},
  {Granato}, {Killedar}, {De Lucia}, {Presotto} \& {Dolag}}{{Cui}
  et~al.}{2014}]{2014MNRAS.437..816C}
{Cui} W.,  {Murante} G.,  {Monaco} P.,  {Borgani} S.,  {Granato} G.~L.,
  {Killedar} M.,  {De Lucia} G.,  {Presotto} V.,    {Dolag} K.,  2014, \mnras,
  437, 816

\bibitem[\protect\citeauthoryear{{Devriendt}, {Guiderdoni} \&
  {Sadat}}{{Devriendt} et~al.}{1999}]{1999A&A...350..381D}
{Devriendt} J.~E.~G.,  {Guiderdoni} B.,    {Sadat} R.,  1999, \aap, 350, 381

\bibitem[\protect\citeauthoryear{{Dubinski}}{{Dubinski}}{1998}]{1998ApJ...502..141D}
{Dubinski} J.,  1998, \apj, 502, 141

\bibitem[\protect\citeauthoryear{{Dubois}, {Devriendt}, {Slyz} \&
  {Teyssier}}{{Dubois} et~al.}{2012}]{2012MNRAS.420.2662D}
{Dubois} Y.,  {Devriendt} J.,  {Slyz} A.,    {Teyssier} R.,  2012, \mnras, 420,
  2662

\bibitem[\protect\citeauthoryear{Eisenstein \& Hu}{Eisenstein \&
  Hu}{1998}]{Eisenstein:1998p1104}
Eisenstein D.~J.,  Hu W.,  1998, Astrophysical Journal v.496, 496, 605

\bibitem[\protect\citeauthoryear{{Emsellem}, {Cappellari}, {Krajnovi{\'c}},
  {Alatalo}, {Blitz}, {Bois}, {Bournaud}, {Bureau}, {Davies}, {Davis}, {de
  Zeeuw} \& {Khochfar}}{{Emsellem} et~al.}{2011}]{2011MNRAS.414..888E}
{Emsellem} E.,  {Cappellari} M.,  {Krajnovi{\'c}} D.,  {Alatalo} K.,  {Blitz}
  L.,  {Bois} M.,  {Bournaud} F.,  {Bureau} M.,  {Davies} R.~L.,  {Davis}
  T.~A.,  {de Zeeuw} P.~T.,    {Khochfar} S. e.~a.,  2011, \mnras, 414, 888

\bibitem[\protect\citeauthoryear{Emsellem, Cappellari, Krajnovi{\'c}, van~de
  Ven, Bacon, Bureau, Davies, de Zeeuw, Falc{\'o}n-Barroso, Kuntschner,
  McDermid, Peletier \& Sarzi}{Emsellem et~al.}{2007}]{Emsellem:2007p5467}
Emsellem E.,  Cappellari M.,  Krajnovi{\'c} D.,  van~de Ven G.,  Bacon R.,
  Bureau M.,  Davies R.~L.,  de Zeeuw P.~T.,  Falc{\'o}n-Barroso J.,
  Kuntschner H.,  McDermid R.,  Peletier R.~F.,    Sarzi M.,  2007, Monthly
  Notices of the Royal Astronomical Society, 379, 401

\bibitem[\protect\citeauthoryear{{Fabian}}{{Fabian}}{2012}]{2012ARA&A..50..455F}
{Fabian} A.~C.,  2012, \araa, 50, 455

\bibitem[\protect\citeauthoryear{Fromang, Hennebelle \& Teyssier}{Fromang
  et~al.}{2006}]{Fromang:2006p400}
Fromang S.,  Hennebelle P.,    Teyssier R.,  2006, Astronomy and Astrophysics,
  457, 371

\bibitem[\protect\citeauthoryear{Gonzalez, Zabludoff \& Zaritsky}{Gonzalez
  et~al.}{2005}]{Gonzalez:2005p907}
Gonzalez A.~H.,  Zabludoff A.~I.,    Zaritsky D.,  2005, The Astrophysical
  Journal, 618, 195

\bibitem[\protect\citeauthoryear{{Gonzalez}, {Zabludoff} \&
  {Zaritsky}}{{Gonzalez} et~al.}{2005}]{2005ApJ...618..195G}
{Gonzalez} A.~H.,  {Zabludoff} A.~I.,    {Zaritsky} D.,  2005, \apj, 618, 195

\bibitem[\protect\citeauthoryear{Gonz{\'a}lez, Audit \& Huynh}{Gonz{\'a}lez
  et~al.}{2007}]{Gonzalez:2007p5292}
Gonz{\'a}lez M.,  Audit E.,    Huynh P.,  2007, Astronomy and Astrophysics,
  464, 429

\bibitem[\protect\citeauthoryear{{Hansen}, {Sheldon}, {Wechsler} \&
  {Koester}}{{Hansen} et~al.}{2009}]{2009ApJ...699.1333H}
{Hansen} S.~M.,  {Sheldon} E.~S.,  {Wechsler} R.~H.,    {Koester} B.~P.,  2009,
  \apj, 699, 1333

\bibitem[\protect\citeauthoryear{{Jimmy}, {Tran}, {Brough}, {Gebhardt}, {von
  der Linden}, {Couch} \& {Sharp}}{{Jimmy} et~al.}{2013}]{2013ApJ...778..171J}
{Jimmy} {Tran} K.-V.,  {Brough} S.,  {Gebhardt} K.,  {von der Linden} A.,
  {Couch} W.~J.,    {Sharp} R.,  2013, \apj, 778, 171

\bibitem[\protect\citeauthoryear{{Kravtsov}, {Vikhlinin} \&
  {Meshscheryakov}}{{Kravtsov} et~al.}{2014}]{2014arXiv1401.7329K}
{Kravtsov} A.,  {Vikhlinin} A.,    {Meshscheryakov} A.,  2014, ArXiv e-prints

\bibitem[\protect\citeauthoryear{{Liu}, {Mao} \& {Meng}}{{Liu}
  et~al.}{2012}]{2012MNRAS.423..422L}
{Liu} F.~S.,  {Mao} S.,    {Meng} X.~M.,  2012, \mnras, 423, 422

\bibitem[\protect\citeauthoryear{Lucia \& Blaizot}{Lucia \&
  Blaizot}{2007}]{DeLucia:2007p1656}
Lucia G.~D.,  Blaizot J.,  2007, Monthly Notices of the Royal Astronomical
  Society, 375, 2

\bibitem[\protect\citeauthoryear{{Martizzi}, {Mohammed}, {Teyssier} \&
  {Moore}}{{Martizzi} et~al.}{2014}]{2014MNRAS.440.2290M}
{Martizzi} D.,  {Mohammed} I.,  {Teyssier} R.,    {Moore} B.,  2014, \mnras,
  440, 2290

\bibitem[\protect\citeauthoryear{{Martizzi}, {Teyssier} \& {Moore}}{{Martizzi}
  et~al.}{2012}]{2012MNRAS.420.2859M}
{Martizzi} D.,  {Teyssier} R.,    {Moore} B.,  2012, \mnras, 420, 2859

\bibitem[\protect\citeauthoryear{{Martizzi}, {Teyssier}, {Moore} \&
  {Wentz}}{{Martizzi} et~al.}{2012}]{2012MNRAS.422.3081M}
{Martizzi} D.,  {Teyssier} R.,  {Moore} B.,    {Wentz} T.,  2012, \mnras, 422,
  3081

\bibitem[\protect\citeauthoryear{{McCarthy}, {Schaye}, {Bower}, {Ponman},
  {Booth}, {Dalla Vecchia} \& {Springel}}{{McCarthy}
  et~al.}{2011}]{2011MNRAS.412.1965M}
{McCarthy} I.~G.,  {Schaye} J.,  {Bower} R.~G.,  {Ponman} T.~J.,  {Booth}
  C.~M.,  {Dalla Vecchia} C.,    {Springel} V.,  2011, \mnras, 412, 1965

\bibitem[\protect\citeauthoryear{{McConnell} \& {Ma}}{{McConnell} \&
  {Ma}}{2013}]{2013ApJ...764..184M}
{McConnell} N.~J.,  {Ma} C.-P.,  2013, \apj, 764, 184

\bibitem[\protect\citeauthoryear{{Mohammed}, {Liesenborgs}, {Saha} \&
  {Williams}}{{Mohammed} et~al.}{2014}]{2014MNRAS.tmp..335M}
{Mohammed} I.,  {Liesenborgs} J.,  {Saha} P.,    {Williams} L.~L.~R.,  2014,
  \mnras

\bibitem[\protect\citeauthoryear{{Moster}, {Naab} \& {White}}{{Moster}
  et~al.}{2013}]{2013MNRAS.428.3121M}
{Moster} B.~P.,  {Naab} T.,    {White} S.~D.~M.,  2013, \mnras, 428, 3121

\bibitem[\protect\citeauthoryear{Moster, Somerville, Maulbetsch, van~den Bosch,
  Macci{\`o}, Naab \& Oser}{Moster et~al.}{2010}]{Moster:2010p5423}
Moster B.~P.,  Somerville R.~S.,  Maulbetsch C.,  van~den Bosch F.~C.,
  Macci{\`o} A.~V.,  Naab T.,    Oser L.,  2010, The Astrophysical Journal,
  710, 903

\bibitem[\protect\citeauthoryear{{Oliva-Altamirano}, {Brough}, {Lidman},
  {Couch}, {Hopkins}, {Colless}, {Taylor}, {Robotham}, {Gunawardhana},
  {Ponman}, {Baldry}, {Bauer} \& {Bland-Hawthorn}}{{Oliva-Altamirano}
  et~al.}{2014}]{2014MNRAS.440..762O}
{Oliva-Altamirano} P.,  {Brough} S.,  {Lidman} C.,  {Couch} W.~J.,  {Hopkins}
  A.~M.,  {Colless} M.,  {Taylor} E.,  {Robotham} A.~S.~G.,  {Gunawardhana}
  M.~L.~P.,  {Ponman} T.,  {Baldry} I.,  {Bauer} A.~E.,    {Bland-Hawthorn} J.
  e.~a.,  2014, \mnras, 440, 762

\bibitem[\protect\citeauthoryear{{Postman}, {Lauer}, {Donahue}, {Graves},
  {Coe}, {Moustakas}, {Koekemoer}, {Bradley}, {Ford}, {Grillo}, {Zitrin},
  {Lemze}, {Broadhurst}, {Moustakas}, {Ascaso}, {Medezinski} \&
  {Kelson}}{{Postman} et~al.}{2012}]{2012ApJ...756..159P}
{Postman} M.,  {Lauer} T.~R.,  {Donahue} M.,  {Graves} G.,  {Coe} D.,
  {Moustakas} J.,  {Koekemoer} A.,  {Bradley} L.,  {Ford} H.~C.,  {Grillo} C.,
  {Zitrin} A.,  {Lemze} D.,  {Broadhurst} T.,  {Moustakas} L.,  {Ascaso} B.,
  {Medezinski} E.,    {Kelson} D.,  2012, \apj, 756, 159

\bibitem[\protect\citeauthoryear{Puchwein, Springel, Sijacki \& Dolag}{Puchwein
  et~al.}{2010}]{Puchwein:2010p763}
Puchwein E.,  Springel V.,  Sijacki D.,    Dolag K.,  2010, eprint arXiv, 1001,
  3018

\bibitem[\protect\citeauthoryear{{Ragone-Figueroa}, {Granato} \&
  {Abadi}}{{Ragone-Figueroa} et~al.}{2012}]{2012MNRAS.423.3243R}
{Ragone-Figueroa} C.,  {Granato} G.~L.,    {Abadi} M.~G.,  2012, \mnras, 423,
  3243

\bibitem[\protect\citeauthoryear{{Ragone-Figueroa}, {Granato}, {Murante},
  {Borgani} \& {Cui}}{{Ragone-Figueroa} et~al.}{2013}]{2013MNRAS.436.1750R}
{Ragone-Figueroa} C.,  {Granato} G.~L.,  {Murante} G.,  {Borgani} S.,    {Cui}
  W.,  2013, \mnras, 436, 1750

\bibitem[\protect\citeauthoryear{{Rudick}, {Mihos}, {Harding}, {Feldmeier},
  {Janowiecki} \& {Morrison}}{{Rudick} et~al.}{2010}]{2010ApJ...720..569R}
{Rudick} C.~S.,  {Mihos} J.~C.,  {Harding} P.,  {Feldmeier} J.~J.,
  {Janowiecki} S.,    {Morrison} H.~L.,  2010, \apj, 720, 569

\bibitem[\protect\citeauthoryear{Sijacki, Springel, Matteo \&
  Hernquist}{Sijacki et~al.}{2007}]{Sijacki:2007p1032}
Sijacki D.,  Springel V.,  Matteo T.~D.,    Hernquist L.,  2007, Monthly
  Notices of the Royal Astronomical Society, 380, 877

\bibitem[\protect\citeauthoryear{Springel, Matteo \& Hernquist}{Springel
  et~al.}{2005}]{Springel:2005p1553}
Springel V.,  Matteo T.~D.,    Hernquist L.,  2005, The Astrophysical Journal,
  620, L79

\bibitem[\protect\citeauthoryear{Stinson, Seth, Katz, Wadsley, Governato \&
  Quinn}{Stinson et~al.}{2006}]{Stinson:2006p1402}
Stinson G.,  Seth A.,  Katz N.,  Wadsley J.,  Governato F.,    Quinn T.,  2006,
  Monthly Notices of the Royal Astronomical Society, 373, 1074

\bibitem[\protect\citeauthoryear{Sutherland \& Dopita}{Sutherland \&
  Dopita}{1993}]{sutherland_dopita93}
Sutherland R.~S.,  Dopita M.~A.,  1993, \apjs, 88, 253

\bibitem[\protect\citeauthoryear{Teyssier}{Teyssier}{2002}]{Teyssier:2002p451}
Teyssier R.,  2002, Astronomy and Astrophysics, 385, 337

\bibitem[\protect\citeauthoryear{Teyssier, Fromang \& Dormy}{Teyssier
  et~al.}{2006}]{Teyssier:2006p413}
Teyssier R.,  Fromang S.,    Dormy E.,  2006, Journal of Computational Physics,
  218, 44

\bibitem[\protect\citeauthoryear{{Teyssier}, {Moore}, {Martizzi}, {Dubois} \&
  {Mayer}}{{Teyssier} et~al.}{2011}]{2011MNRAS.414..195T}
{Teyssier} R.,  {Moore} B.,  {Martizzi} D.,  {Dubois} Y.,    {Mayer} L.,  2011,
  \mnras, 414, 195

\bibitem[\protect\citeauthoryear{Toro, Spruce \& Speares}{Toro
  et~al.}{1994}]{Toro:1994p1151}
Toro E.~F.,  Spruce M.,    Speares W.,  1994, Shock Waves, 4, 25

\bibitem[\protect\citeauthoryear{{Tweed}, {Devriendt}, {Blaizot}, {Colombi} \&
  {Slyz}}{{Tweed} et~al.}{2009}]{2009A&A...506..647T}
{Tweed} D.,  {Devriendt} J.,  {Blaizot} J.,  {Colombi} S.,    {Slyz} A.,  2009,
  \aap, 506, 647

\bibitem[\protect\citeauthoryear{{van den Bosch}, {Yang} \& {Mo}}{{van den
  Bosch} et~al.}{2003}]{2003MNRAS.340..771V}
{van den Bosch} F.~C.,  {Yang} X.,    {Mo} H.~J.,  2003, \mnras, 340, 771

\bibitem[\protect\citeauthoryear{van~der Wel, Holden, Zirm, Franx, Rettura,
  Illingworth \& Ford}{van~der Wel et~al.}{2008}]{vanderWel:2008p5391}
van~der Wel A.,  Holden B.~P.,  Zirm A.~W.,  Franx M.,  Rettura A.,
  Illingworth G.~D.,    Ford H.~C.,  2008, The Astrophysical Journal, 688, 48

\bibitem[\protect\citeauthoryear{{von der Linden}, {Best}, {Kauffmann} \&
  {White}}{{von der Linden} et~al.}{2007}]{2007MNRAS.379..867V}
{von der Linden} A.,  {Best} P.~N.,  {Kauffmann} G.,    {White} S.~D.~M.,
  2007, \mnras, 379, 867

\end{thebibliography}

%%%%%%%%%%%%%%%%%%%%%%%%%%%%%%%%%%%%%%%%%%%%%%%%%%%%%%%%%%%

\appendix
\section{Stellar mass vs. halo mass with Sersic fits}\label{appendix:A}

\begin{figure}
    \includegraphics[width=0.49\textwidth]{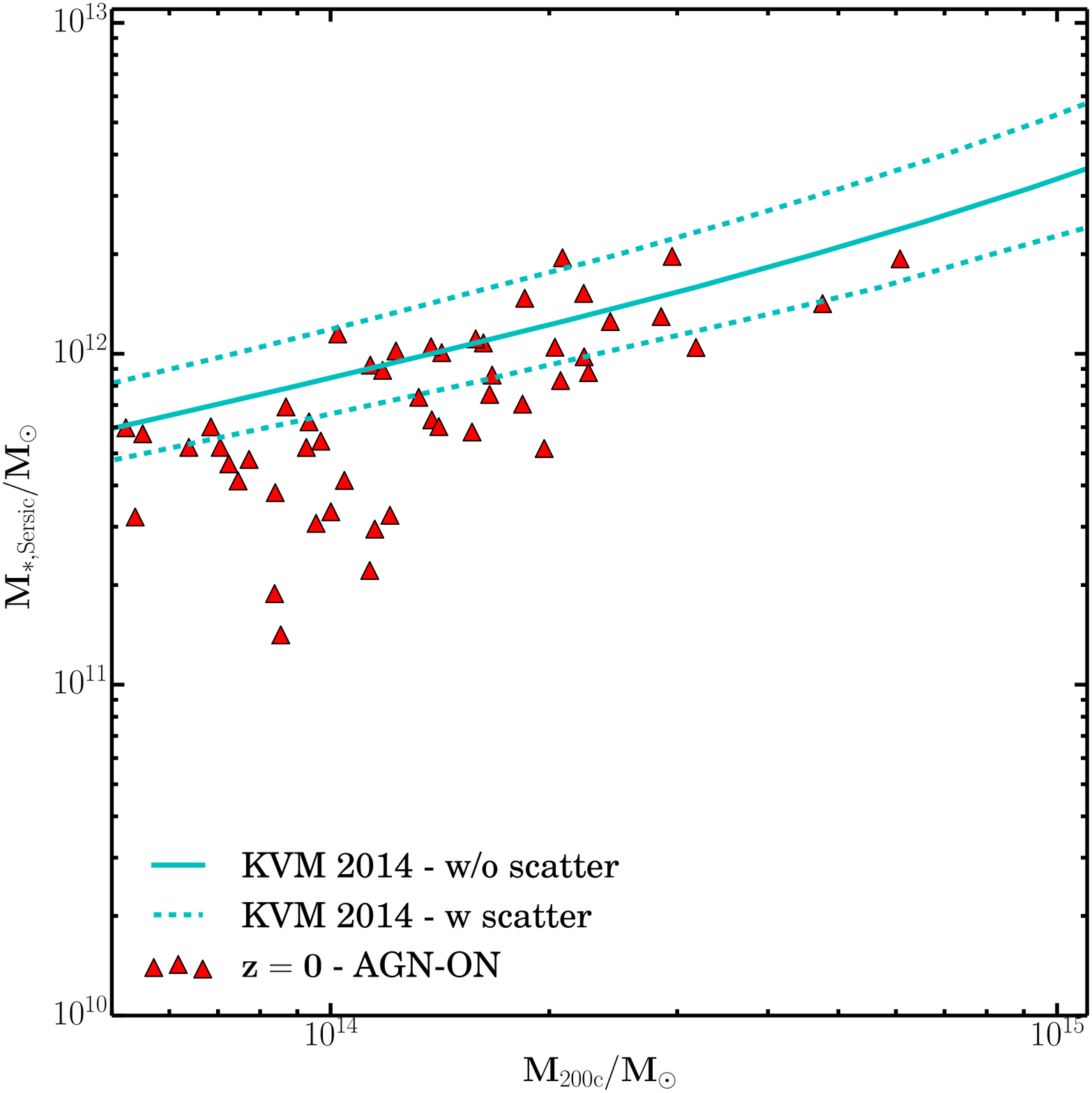}
\caption{Results from our simulations versus the halo mass vs. stellar mass relation at redshift $z=0$ from KVM 2014. Our AGN-ON simulations are represented 
by the red triangles. The BCG stellar mass definition is based on double Sersic fits.}
  \label{fig:abmatchsersic}
\end{figure}

We have tried several definitions for the stellar mass we plot in Figure~\ref{fig:abmatch}. In this Appendix we show the results we obtain when we adopt an alternative mass definition with respect to the one shown in the main text. 
Since the differences are not huge we only show the results here. A straightforward approach to obtain stellar masses from simulations is to fit the surface brightness profiles with an analytical function. After excluding the satellite galaxies from the analysis, we fit a double Sersic profile (the sum of two Service profiles) to the BCG+ICL component. We find that a double Sersic function is generally an excellent fit to the simulated data. We associate the BCG to the inner stellar component, we estimate its stellar mass and we plot it against the halo mass in Figure~\ref{fig:abmatchsersic}. For the most massive halos, the stellar masses obtained with this procedure are slightly larger than the ones obtained by adopting the surface brightness cut, however the difference is somewhat smaller for the less massive halos. 

%%%%%%%%%%%%%%%%%%%%%%%%%%%%%%%%%%%%%%%%%%%%%%%%%%%%%%%%%%%

\label{lastpage}
\end{document}